\documentclass[%
 reprint,
 superscriptaddress,
 amsmath,amssymb,
 aps,
]{revtex4-2}

\usepackage{graphicx}
\usepackage{dcolumn}
\usepackage{bm}
\usepackage{physics} 
\usepackage{braket}
\usepackage{hyperref}
\hypersetup{
  colorlinks   = true,    
  urlcolor     = blue,    
  linkcolor    = blue,    
  citecolor    = blue      
}
\usepackage[capitalize]{cleveref}
\usepackage{accents}
\usepackage{enumerate}
\usepackage[T1]{fontenc}

\DeclareDocumentCommand\ceil{ s m }
{
    \lceil #2 \rceil
}
\DeclareDocumentCommand\floor{ s m }
{
    \lfloor #2 \rfloor
}

\usepackage{xcolor}
\definecolor{myyellow}{HTML}{f1a226}
\definecolor{mygreen}{HTML}{298c8c}
\usepackage{graphicx}
\usepackage{placeins}

\begin{document}

\preprint{APS/123-QED}

\title{Single-Qubit Gates Beyond the Rotating-Wave Approximation \\for Strongly Anharmonic Low-Frequency Qubits}

\author{Martijn F. S. Zwanenburg}
\email{m.f.s.zwanenburg@tudelft.nl}
\affiliation{QuTech and Kavli Institute of Nanoscience, Delft University of Technology, 2628 CJ, Delft, The Netherlands}
\author{Siddharth Singh}
\affiliation{QuTech and Kavli Institute of Nanoscience, Delft University of Technology, 2628 CJ, Delft, The Netherlands}
\author{Eugene Y. Huang}
\affiliation{QuTech and Kavli Institute of Nanoscience, Delft University of Technology, 2628 CJ, Delft, The Netherlands}
\author{Figen Yilmaz}
\affiliation{QuTech and Kavli Institute of Nanoscience, Delft University of Technology, 2628 CJ, Delft, The Netherlands}
\author{\mbox{Taryn V. Stefanski}}
\affiliation{QuTech and Kavli Institute of Nanoscience, Delft University of Technology, 2628 CJ, Delft, The Netherlands}
\affiliation{\mbox{Quantum Engineering Centre for Doctoral Training, H. H. Wills Physics Laboratory} and Department of Electrical and Electronic Engineering, University of Bristol, BS8 1FD, Bristol, UK}
\author{Jinlun Hu}
\affiliation{QuTech and Kavli Institute of Nanoscience, Delft University of Technology, 2628 CJ, Delft, The Netherlands}
\author{Piranavan Kumaravadivel}
\affiliation{QuTech and Netherlands Organization for Applied Scientific Research (TNO), 2628 CJ, Delft, The Netherlands}
\author{Christian Kraglund Andersen}
\email{c.k.andersen@tudelft.nl}
\affiliation{QuTech and Kavli Institute of Nanoscience, Delft University of Technology, 2628 CJ, Delft, The Netherlands}

\date{\today}

\begin{abstract}
Single-qubit gates are in many quantum platforms applied using a linear drive resonant with the qubit transition frequency which is often theoretically described within the rotating-wave approximation (RWA). However, for fast gates on low-frequency qubits, the RWA may not hold and we need to consider the contribution from counter-rotating terms to the qubit dynamics. The inclusion of counter-rotating terms into the theoretical description gives rise to two challenges. Firstly, it becomes challenging to analytically calculate the time evolution as the Hamiltonian is no longer self-commuting. Moreover, the time evolution now depends on the carrier phase such that, in general, every operation in a sequence of gates is different. In this work, we derive and verify a correction to the drive pulses that minimizes the effect of these counter-rotating terms in a two-level system. We then derive a second correction term that arises from non-computational levels for a strongly anharmonic system. We experimentally implement these correction terms on a fluxonium superconducting qubit, which is an example of a strongly anharmonic, low-frequency qubit for which the RWA may not hold, and demonstrate how fast, high-fidelity single-qubit gates can be achieved without the need for additional hardware complexities. 
\end{abstract}

\maketitle


\section{\label{sec:introduction}Introduction}
Single-qubit operations lie at the heart of many potential applications of quantum computing \cite{nielsen-chuang, preskill_nisq_era}. In many quantum computing architectures, such as NV \mbox{centers \cite{nv_center1}}, quantum dots \cite{qd1, qd2}, trapped ions \cite{trapped_ions1, trapped_ion2}, and superconducting circuits \cite{scq_review_1,scq_review_2}, these operations are implemented using linearly polarized signals. This linear drive signal consists of two components: one that rotates in the same direction as the qubit on the Bloch sphere and one that rotates in the opposite direction, which are denoted as the co-rotating and counter-rotating terms respectively. The theoretical description of these operations often relies on making the rotating-wave approximation (RWA), in which the counter-rotating terms are dropped resulting in a Hamiltonian that is exactly solvable when the drive strength is time-independent or when the drive frequency is on resonance with the qubit transition frequency \cite{rabi1,rabi2}. Making the RWA is justified when the ratio between the drive strength and the drive frequency is small. There are two dominant error channels when considering the implementation of such a drive signal on a realistic system: the shift in qubit frequency known as the ac Stark shift and leakage to levels outside the computational subspace \cite{ac_stark,drag_error_terms}. Suppressing these two error channels can be achieved by using pulse-shaping techniques \cite{drag1, drag2, wahah, leakage_characterization, leakage_characterization2, pwc-pulses, recursive-drag, hd-drag}. 

When the ratio between the drive strength and the drive frequency is increased and the RWA becomes invalid, the dynamics of the system becomes richer in nature \cite{bloch_siegert, breaking_rwa_single_electron_spin, breaking_rwa_lasers, circular_polarized_drive, operations_qsl, nv_beyond_rwa}. For example, the time evolution now depends on the phase of the carrier signal such that it depends on the absolute time at which the drive signal is applied. Furthermore, there is no straightforward way to analytically derive the time evolution, making it challenging to compute the pulse parameters that maximize the gate fidelity.

Recently, Ref. \cite{circular_polarized_drive} proposed and implemented two solutions to these challenges based on circularly polarized driving and commensurate gate durations. However, these solutions require additional hardware complexities and do not take non-computational levels into account. In this work, we show analytically, numerically and experimentally that it is possible to achieve low gate errors by using conventional hardware and pulse shaping techniques. We employ the Magnus-Taylor expansion introduced in \mbox{Ref. \cite{exact_rwa}} to derive optimal pulse shapes for a two-level system undergoing a drive beyond the RWA. In addition to computing these optimal pulse shapes for a two-level system, we derive and verify an additional correction term arising from non-computational levels for a strongly anharmonic, low-frequency system. Finally, we use the derived pulse shapes and the understanding of non-RWA dynamics developed in this work to formulate deterministic experimental calibration protocols. Numerically we find that, using relatively straightforward calibration procedures and for the system studied in this work, it is possible to achieve coherent error rates averaged over $\pi$ and $\pi/2$ gates below $10^{-6}$ for gate durations as short as 2.64 Larmor periods.

\begin{figure}[t]
\centering

\includegraphics[width=0.5\linewidth]{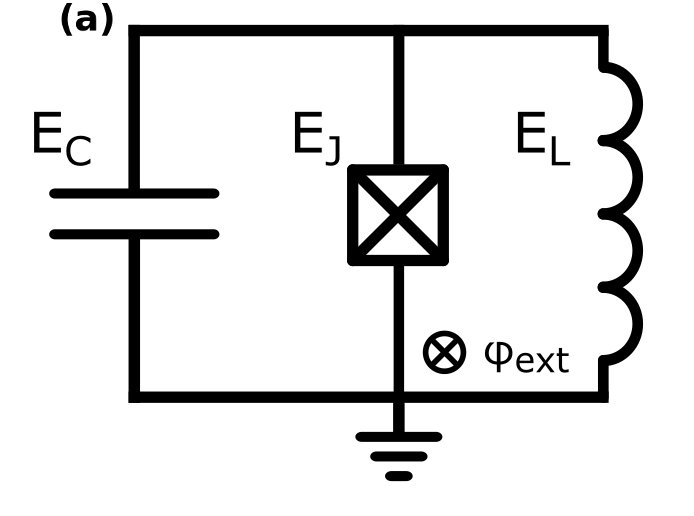}%
\includegraphics[width=0.5\linewidth]{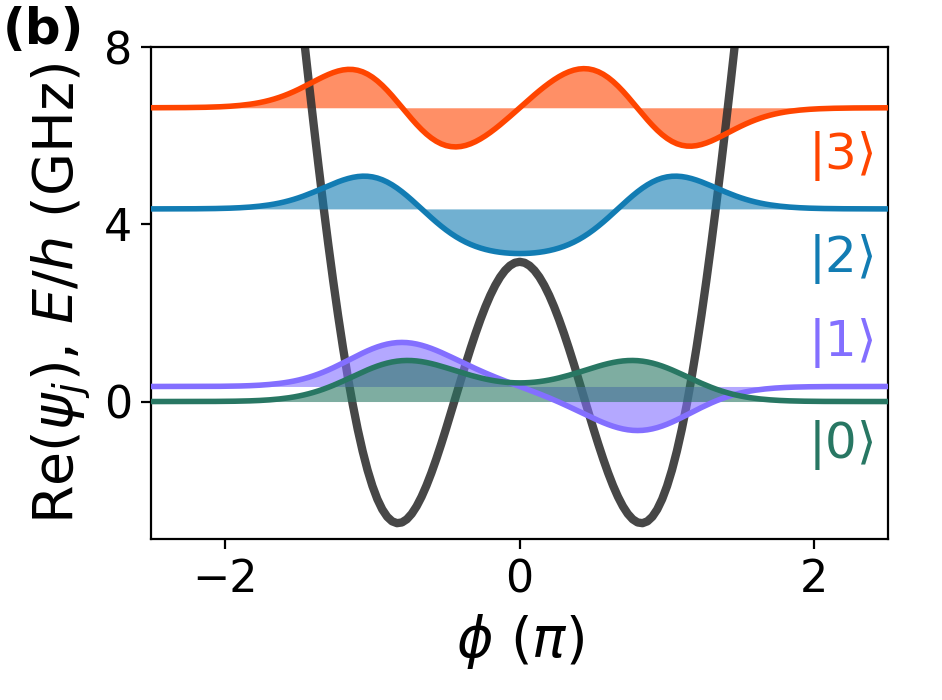}

\caption{\label{fig:figure4} (a) Circuit diagram of the fluxonium qubit. \mbox{(b) Potential} and the wavefunctions offset by the eigenenergies of the first four levels for $E_C/h=E_L/h=$ 1 GHz and $E_J/h=$ 5 GHz at the sweet spot $\varphi_\text{ext}=0.5$.}
\end{figure}

While the general theory in this work is applicable to any strongly anharmonic, low-frequency system, we are specifically inspired by the fluxonium qubit as one such example \cite{fluxonium}. A fluxonium qubit consists of a capacitor shunted by a single Josephson junction and a linear inductor characterized by the energies $E_C$, $E_J$ and $E_L$ respectively. The circuit diagram of a fluxonium qubit is shown in Fig. \ref{fig:figure4}(a), and the Hamiltonian of such a circuit is given by:

\begin{equation}
    \hat{H} = 4E_C\hat{n}^2 + \frac{1}{2}E_L\hat{\phi}^2 - E_J\cos(\hat{\phi} - 2\pi\varphi_\text{ext}).
\end{equation}

\noindent Here, $\hat{n}$ and $\hat{\phi}$ are the number-of-Cooper-pairs and phase operators and $\varphi_\text{ext}=\Phi_\text{ext}/\Phi_0$ is the reduced external flux with $\Phi_\text{ext}$ the external flux and $\Phi_0$ the magnetic flux quantum. When the fluxonium is tuned to its sweet-spot at $\varphi_\text{ext}=0.5$, the potential mimics a double-well potential, where the wavefunctions of the $\ket{0}$ and $\ket{1}$ states are symmetric and anti-symmetric superpositions of states living in the left and right well. The potential and wavefunctions for a fluxonium with example parameters $E_C/h=E_L/h=$ 1 GHz and $E_J/h=$ 5 GHz are computed using \texttt{scqubits} and shown in Fig. \ref{fig:figure4}(b) \cite{scqubits1,scqubits2}. The qubit transition frequency of a typical fluxonium at its operating point is in the $100-1000$ MHz range \cite{microwave_activated_cz, circular_polarized_drive, high_coherence_fluxonium, fluxonium_cnot, fluxonium_tunable_coupler, fluxonium_two_qubit_gate_alibaba, fluxonium_review_berkeley, millisecond_coherence, coupler_activated_cphase, fluxonium_cross_resonance, mit_cz, fluxonium_initialization, fpa_experimental, EPR}, but can also be in the $1-100$ MHz range by designing a heavier fluxonium \cite{fast_logic_slow_qubits, ac_charge_detection, tunable_inductive_coupling}. Since single-qubit gate durations on fluxonium qubits are typically in the 10-100 ns range, resulting in drive strengths in the 10-100 MHz range, they are an ideal platform to investigate non-RWA effects in single-qubit operations.

\section{\label{sec:theory} Theory: Single-Qubit Gates Beyond the RWA}
Our description of single-qubit gates starts with a general drive Hamiltonian for a two-level system \cite{cqed}:

\begin{equation}
\label{eq:two-level-hamiltonian}
    H/\hbar = \omega_q \ket{1}\bra{1} + \mathcal{D}(t) \big(\ket{0}\bra{1}+\ket{1}\bra{0}\big),
\end{equation}

\noindent where $\omega_q$ is the qubit frequency and $\mathcal{D}(t)$ is the drive signal defined by: 

\begin{equation}
\label{eq:drive_term_definition}
\begin{split}
    \mathcal{D}(t) &= \Re(\mathcal{E}(t))\cos(\omega_dt+\phi) + \Im(\mathcal{E}(t))\sin(\omega_dt+\phi), \\
    \mathcal{E}(t) &= e^{i\theta}(\mathcal{E}_I(t)+i\mathcal{E}_Q(t)).
\end{split}
\end{equation}

\noindent Here, $\omega_d$ is the drive frequency, $\phi$ the phase of the carrier signal, $\theta$ the angle of the rotation axis w.r.t. the $X$-axis of the Bloch sphere and $\mathcal{E}_I(t)$ and $\mathcal{E}_Q(t)$ the in-phase and quadrature pulse envelopes respectively. For now, we restrict ourselves to $X$ rotations by setting $\theta=0$. We proceed to transform the Hamiltonian into the rotating frame $\Tilde{H}(t) = V(t)H(t)V^\dagger(t) + i \dot{V}(t)V^\dagger(t)$ with $V(t)=\exp{i(\omega_dt + \phi)\ket{1}\bra{1}}$ to obtain:

\begin{equation}
\label{eq:hamiltonian-terms}
\begin{split}
    &\tilde{H}(t)/\hbar = -\frac{\Delta}{2}\sigma_z + A_I(t)\sigma_x + A_Q(t)\sigma_y, \\
    &A_I(t) = \frac{1}{2}\Big(\mathcal{E}_I(t)\left(1+\cos(2\omega_dt+2\phi)\right) + \\
    & \hspace{2.48cm} \mathcal{E}_Q(t)\sin(2\omega_dt+2\phi)\Big), \\
    &A_Q(t) = \frac{1}{2}\Big(\mathcal{E}_Q(t)\left(1-\cos(2\omega_dt+2\phi)\right) + \\
    & \hspace{2.7cm} \mathcal{E}_I(t)\sin(2\omega_dt+2\phi)\Big),
\end{split}
\end{equation}

\noindent with $\Delta=\omega_q-\omega_d$ the detuning between the qubit and drive frequency. The in-phase and quadrature components of the drive signal both contribute to one stationary term and two non-stationary terms that oscillate at $2\omega_d$. If we invoke the RWA and drop the oscillating terms we arrive at the RWA Hamiltonian:

\begin{equation}
\label{eq:hamiltonian-terms-rwa}
    \tilde{H}_\text{RWA}(t)/\hbar = -\frac{\Delta}{2}\sigma_z + \frac{1}{2}\big(\mathcal{E}_I(t)\sigma_x + \mathcal{E}_Q(t)\sigma_y\big).
\end{equation}

Here, we are interested in determining the pulse parameters that implement a given operation with maximum fidelity. Specifically, we consider an operation applied from $t=0$ to $t=t_g$, with $t_g$ the gate duration. If the RWA holds, it is straightforward to derive the pulse parameters that implement any arbitrary rotation on the Bloch sphere. For example, for an $X_\pi$ gate we set $\Delta=\mathcal{E}_Q=0$ and $\int_0^{t_g} dt \mathcal{E}_{I}(t) = \pi$. In contrast, if the RWA does not hold, there are two factors that significantly complicate the description of single-qubit gates. Firstly, the time evolution is no longer independent of the carrier phase $\phi$, such that the time evolution depends on the absolute time at which the gate is applied. Secondly, the non-RWA Hamiltonian $\tilde{H}$ is not self-commuting such that an analytical expression for the time evolution is challenging to obtain. Even if an (approximate) expression can be computed it may be challenging to derive the ideal pulse parameters from that solution.

To resolve these challenges we make use of the methods introduced in Ref. \cite{exact_rwa}. Rather than computing the time evolution over the full duration of the operation, the time interval on which the time evolution is calculated is restricted to Magnus intervals $t_{m,n} = [t_0+(n-m)t_c,t_0+nt_c]$. Here, $t_c=\pi/\omega_d$ is the Magnus period, which is the period of the non-stationary terms in the Hamiltonian, and $m$ and $n$ are positive integers such that the time interval $t_{m,n}$ contains $n-m$ Magnus periods. As will be shown, it is possible to derive corrections for the non-RWA terms in the Hamiltonian by restricting to the Magnus intervals $t_{m,n}$. To derive these corrections, the key idea is to intersect the non-RWA and RWA time evolutions on these Magnus intervals:

\begin{equation}
\label{eq:fundamental-equation}
\begin{split}
    &\mathcal{T}\exp(-\frac{i}{\hbar}\int_{t_0+(n-m)t_c}^{t_0+nt_c}d\tau \tilde{H}_\text{RWA}(\tau,\mathcal{P}_\text{RWA})) \\
    = &\mathcal{T}\exp(-\frac{i}{\hbar}\int_{t_0+(n-m)t_c}^{t_0+nt_c}d\tau \tilde{H}(\tau,\mathcal{P})),
\end{split}
\end{equation}

\noindent with $\mathcal{T}$ the time-ordering operator and the dependence of both Hamiltonians on the pulse parameters is shown explicitly. $\mathcal{P}_\text{RWA}$ are the ideal pulse parameters if the RWA holds exactly and $\mathcal{P}$ are the to-be-derived pulse parameters that intersect the non-RWA and RWA time evolutions on the specified interval. Notice that, unless the gate is commensurate (i.e., the gate duration is an exact integer multiple of the Magnus period), \mbox{Eq. \eqref{eq:fundamental-equation}} can not capture the full evolution from $t=0$ to $t=t_g$. However, we can maximize the interval on which the RWA and non-RWA evolutions are equivalent by satisfying \mbox{Eq. \eqref{eq:fundamental-equation}} from $t=t_0$ to $t=t_0+N_ct_c$ with $N_c=\floor{\frac{t_g-t_0}{t_c}}$ the number of Magnus periods during the operation where $\floor{.}$ is the floor function. If \mbox{Eq. \eqref{eq:fundamental-equation}} is satisfied, the non-RWA Hamiltonian implements the same time evolution $U(t_0,t_0+N_ct_c)$ as the RWA Hamiltonian. Since the RWA Hamiltonian implements the desired operation $U(0,t_g)$ with unit fidelity, we expect the pulse parameters $\mathcal{P}$ that solve \mbox{Eq. \eqref{eq:fundamental-equation}} to also minimize the error in the full non-RWA time evolution from $t=0$ to $t=t_g$. This is based on the assumption that the effect of the non-RWA terms is negligible on the uncorrected time intervals with $t \in [0,t_0] \cup [N_ct_c+t_0,t_g]$. We expect this to be the case for sufficiently long gates as the uncorrected intervals are then small compared to the total time evolution. Consequently, we expect this framework to break down as $N_c$ approaches $1$.

There are two choices for $n$ and $m$ for which Eq. \eqref{eq:fundamental-equation} is satisfied on the interval $t \in [t_0,t_0+N_ct_c]$. We can either satisfy Eq. \eqref{eq:fundamental-equation} for all $n$ and $m=1$ such that it is satisfied for all Magnus intervals during the gate. The other choice is to directly integrate over $N_c$ Magnus periods by setting $m=n=N_c$. In this work we focus on the latter.

To derive the pulse parameters $\mathcal{P}$ that satisfy \mbox{Eq. \eqref{eq:fundamental-equation}} we Magnus expand the non-RWA time evolution, of which the first two terms are given by \cite{magnus_expansion, magnus_review}:

\begin{equation}
\label{eq:magnus-definition-1}
\begin{split}
    & \overline{H}^{(0)} = \int_{t_0 + (n-m)t_c}^{t_0+nt_c}dt_1 H(t_1), \\
    & \overline{H}^{(1)} = -\frac{i}{2}\int_{t_0 + (n-m)t_c}^{t_0+nt_c}dt_1\int_{t_0}^{t_1}dt_2 \Big[H(t_1),H(t_2)\Big],
\end{split}
\end{equation}

\noindent such that we can approximate the time evolution over the Magnus periods $t_{m,n}$ by:

\begin{equation}
\label{eq:magnus-definition-2}
\begin{split}
    U(t_0+t_c,t_0) &= \mathcal{T}\exp(-\frac{i}{\hbar}\int_{t_0 + (n-m)t_c}^{t_0+nt_c}d\tau H(\tau)) \\
    &= \exp(-i\overline{H}/\hbar), \quad \overline{H} = \sum_{k=0}^\infty \overline{H}^{(k)}.
\end{split}
\end{equation}

\noindent The time-ordering operator for the RWA time evolution can typically be dropped as the RWA Hamiltonian $\tilde{H}_\text{RWA}$ commutes with itself for $\Delta=\mathcal{E}_Q=0$ (in an appropriate frame). Using \cref{eq:hamiltonian-terms,eq:hamiltonian-terms-rwa,eq:fundamental-equation,eq:magnus-definition-1,eq:magnus-definition-2} it is possible to derive expressions for the ideal pulse parameters. In the following sections, we will derive and numerically verify these expressions for the zeroth-order and first-order Magnus expansion terms which we will refer to as the zeroth/first-order Magnus approximation. 

\subsection{Zeroth-Order Magnus Expansion \label{sec:theory-0th-order}}
While the formalism derived here is applicable to any smooth pulse envelope, it is convenient to explicitly define the pulse envelope $\mathcal{E}_I(t)$. In this work we focus specifically on the cosine pulse envelope defined as:

\begin{equation}
\label{eq:pulse-shape-definitions}
\begin{split}
    \mathcal{E}_I(t) &= \Omega_Is_I(t) = \frac{\Omega_I}{2}\left(1-\cos(\frac{2\pi}{t_g}t)\right),
\end{split}
\end{equation}

\noindent with $\Omega_I$ the drive strength. Using \cref{eq:hamiltonian-terms,eq:hamiltonian-terms-rwa,eq:fundamental-equation,eq:magnus-definition-1,eq:magnus-definition-2} we obtain three requirements for the pulse parameters in the zeroth-order Magnus approximation:

\begin{equation}
\label{eq:magnus0-three-requirements}
\begin{split}
    &(\text{i}) \;\; \int_{b_-}^{b_+}dt_1 \frac{-\Delta}{2} = 0, \\
    &(\text{ii}) \;\; \int_{b_-}^{b_+}dt_1 A_Q(t_1,\mathcal{P}) = 0, \\
    &(\text{iii}) \;\; \int_{b_-}^{b_+}dt_1 A_I(t_1,\mathcal{P}) = \int_{b_-}^{b_+}dt_1 \frac{\mathcal{E}_{I}(t_1,\mathcal{P}_\text{RWA})}{2},
\end{split}
\end{equation}

\noindent where we have substituted the general bounds by $b_-=t_0+(n-m)t_c$ and $b_+=t_0+nt_c$. Requirement (i) is straightforwardly satisfied by setting $\Delta=0$. By Taylor expanding the pulse envelopes we find that we can satisfy requirement (ii) by defining the quadrature pulse envelope as:

\begin{equation}
\label{eq:pulse-shape-definitions2}
    \mathcal{E}_Q(t) = \lambda\frac{\partial \mathcal{E}_I(t)}{\partial t} = \lambda \Omega_I \frac{\pi}{t_g}\sin(\frac{2\pi}{t_g}t),
\end{equation}

\noindent with $\lambda$ defining the relative drive strength of the quadrature drive signal which we will refer to as the pulse proportionality parameter (PPP). A full derivation for the quadrature pulse envelope is provided in App. \ref{app:zeroth-order-magnus-expansion}. We find that requirement (iii) can be satisfied independently of the shape of the quadrature pulse envelope. By computing the integrals in requirements (ii) and (iii) we obtain closed-form expressions for the PPP $\lambda$ and the drive strength $\Omega_I$:

\begin{widetext}
\begin{subequations}
\label{eq:magnus0-requirement}

\begin{equation}
\label{eq:magnus0-lambda-requirement}
    \lambda = \frac{\sum_{k=0}^\infty \left(\frac{1}{2\omega_d}\right)^{k+1} \gamma_2(k,\beta) \left[\frac{\partial^{k}s_I(t_1)}{\partial t_1^{k}}\right]_{t_1=b_-}^{b_+}}{s_I(b_+) - s_I(b_-) - \sum_{k=0}^\infty \left(\frac{1}{2\omega_d}\right)^{k+1} \gamma_1(k,\beta) \left[\frac{\partial^{k+1}s_I(t_1)}{\partial t_1^{k+1}}\right]_{t_1=b_-}^{b_+}},
\end{equation}

\begin{equation}
\label{eq:magnus0-omega-requirement}
    \frac{\Omega_I}{\Omega_{I,\text{RWA}}} = \frac{\int_{b_-}^{b_+}dt_1s_I(t_1)}{ \int_{b_-}^{b_+}dt_1s_I(t_1) + \sum_{k=0}^\infty \left(\frac{1}{2\omega_d}\right)^{k+1} \Bigg\{ \gamma_1(k,\beta) \left[\frac{\partial^{k}s_I(t_1)}{\partial t_1^{k}}\right]_{t_1=b_-}^{b_+} - \lambda \gamma_2(k,\beta) \left[\frac{\partial^{k+1}s_I(t_1)}{\partial t_1^{k+1}}\right]_{t_1=b_-}^{b_+} \Bigg\}}.
\end{equation}

\end{subequations}
\end{widetext}

\noindent Here, $\Omega_{I,\text{RWA}}$ is the ideal drive strength for the RWA Hamiltonian and we have defined:

\begin{equation}
\label{eq:gamma-pm-def}
\begin{split}
    \gamma_1(k,\beta) &= (-1)^{\floor{\frac{k}{2}}} \Big(\chi_+(k)\sin(\beta) + \chi_-(k)\cos(\beta)\Big), \\
    \gamma_2(k,\beta) &= (-1)^{\floor{\frac{k+1}{2}}} \Big(\chi_+(k)\cos(\beta) + \chi_-(k)\sin(\beta)\Big),
\end{split}
\end{equation}

\noindent with $\beta=2\omega_dt_0+2\phi$ and $\chi_\pm(k) = \frac{1\pm(-1)^k}{2}$. As apparent from Eq. \eqref{eq:magnus0-requirement}, the expressions for the pulse parameters are given by an infinite series. The magnitude of the terms in these sums scale with $(\pi/\omega_dt_g)^{k}=(t_c/t_g)^{k}$, since $\frac{\partial^ks_I(t)}{\partial t^k}\sim (2\pi/t_g)^k$. As we are specifically interested in operations for which the drive strength approaches the same order of magnitude as the drive frequency, i.e. $1 \lesssim t_g/t_c \lesssim 10$, convergence of these infinite series might be slow, so it is important to truncate these sums at sufficiently high $k$. For $t_g<t_c$ the ideal pulse parameters calculated in Eq. \eqref{eq:magnus0-requirement} are undefined since there are no Magnus intervals to integrate over in Eq. \eqref{eq:fundamental-equation}.

As apparent from Eqs. \eqref{eq:magnus0-requirement} and \eqref{eq:gamma-pm-def}, the ideal pulse parameters depend on the carrier phase $\phi$ through the parameter $\beta$. This is problematic, since every gate in a sequence of operations generally has a different starting carrier phase, and varying the pulse parameters for each gate is highly impractical. Therefore, the ideal pulse parameters have to be defined as an average of the expressions in Eq. \eqref{eq:magnus0-requirement} over the carrier phase $\phi$. From this, we intuitively understand how the breakdown of the RWA causes the average gate error to increase: when the RWA becomes increasingly invalid, the dependence of the time evolution on the carrier phase increases, which increases the variance of the pulse parameters calculated in \mbox{Eq. \eqref{eq:magnus0-requirement}} such that the pulse parameters averaged over the carrier phase no longer result in low gate errors for all carrier phases. The pulse parameters also depend on the start of the integration window $t_0$. This parameter, unlike the carrier phase, does not represent a physical parameter from the drive pulse. Instead, we are free to choose $t_0$ in \mbox{Eq. \eqref{eq:magnus0-requirement}}, where this choice is represented through the parameters $\beta$ and $b_\pm$. This indicates that for each carrier phase $\phi$ there may exist a range of ideal pulse parameters corresponding to different choices of $t_0$ rather than one unique solution. 

Two open questions remain: can the dependence of the ideal pulse parameters on the carrier phase $\phi$ be eliminated, and how do we choose $\beta$ and $b_\pm$ or equivalently the start of the integration window $t_0$? As detailed in App. \ref{app:beta0}, we can leverage the freedom to choose these parameters to derive exact algebraic solutions for $\Omega_I$ and $\lambda$ in \mbox{Eq. \eqref{eq:magnus0-requirement}}. Strikingly, we find that, by simply setting $\lambda = 1/2\omega_d$ and $\Omega_I=\Omega_{I,\text{RWA}}$, the non-RWA terms in the zeroth order Magnus approximation are corrected independently of the carrier phase and over the full time evolution from $t=0$ to $t=t_g$ rather than only over Magnus intervals. This solution corresponds to setting $\beta=0$ or by using a symmetric integration window around $t=t_g/2$ such that $t_0=(t_g-N_ct_c)/2$. We highlight that these solutions exist as long as the pulse envelopes are analytic on the interval $t \in [0,t_g]$ and if $s_I(0)=s_I(t_g)=0$. For the symmetric integration windows it is additionally required that each even (odd) \mbox{$k$-th} time-derivative of $s_I(t)$ is symmetric (anti-symmetric) around $t=t_g/2$. We emphasize that these solutions only exist for the specified parameters, and that we have not been able to derive similar solutions in higher-order Magnus approximations. Therefore, the algebraic solution is only used in the zeroth-order Magnus approximation and for the specified parameters, and in all other situations the terms in \mbox{Eq. \eqref{eq:magnus0-requirement}} have to be computed directly, which we will refer to as the truncated series solution.

\begin{figure}[t]
\centering
\includegraphics[width=\linewidth]{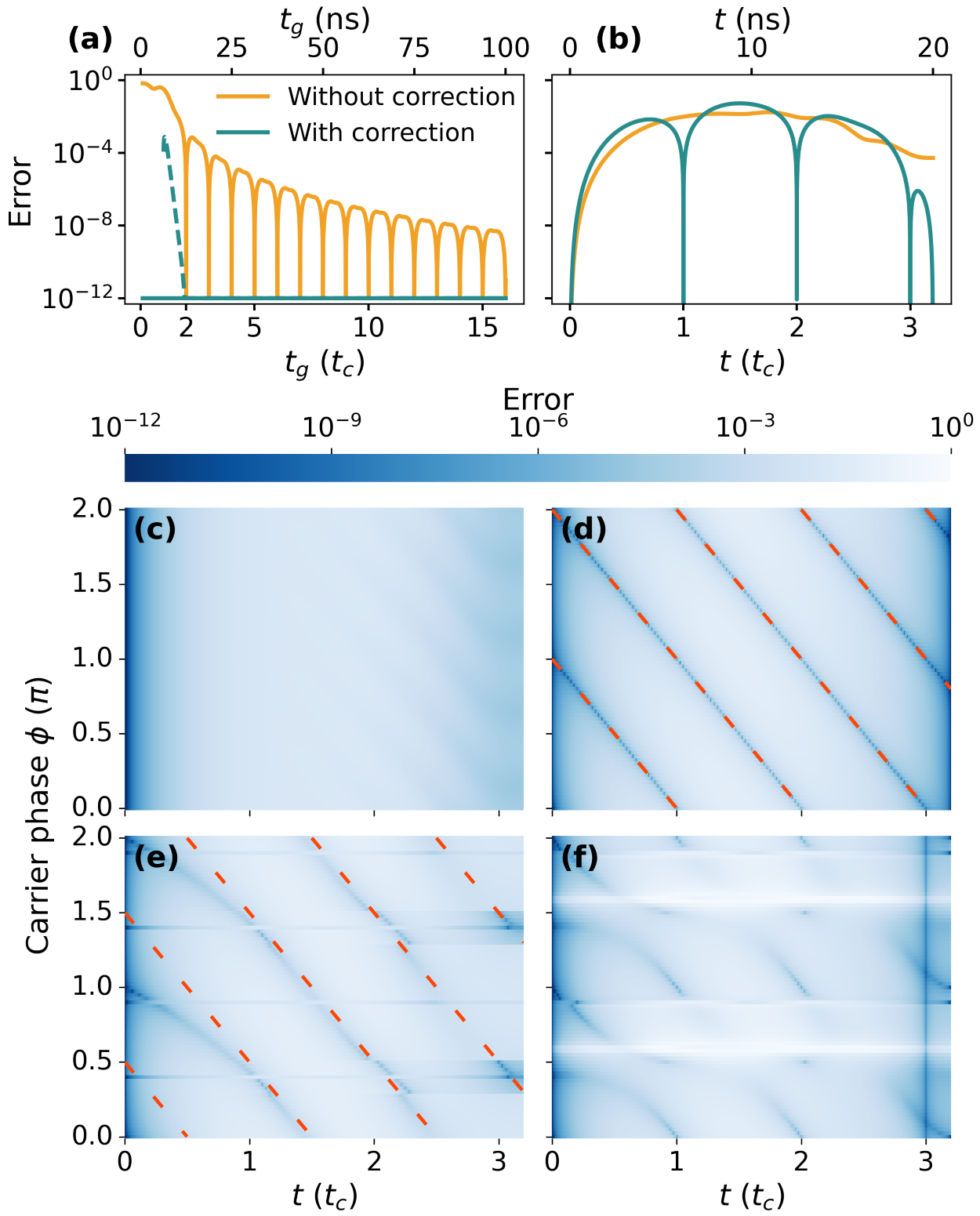}
\caption{\label{fig:figure12} Error between the zeroth-order Magnus expanded non-RWA time evolution and the desired RWA time evolution for $\omega_{01}/2\pi = \omega_d/2\pi = 80$ MHz. Errors are truncated at $10^{-12}$ for visibility purposes. (a) Gate-error as a function of the gate duration $t_g$. The uncorrected pulse parameters correspond to setting $\mathcal{P}=\mathcal{P}_\text{RWA}$. The solid corrected line corresponds to the algebraic result, i.e. $\Omega_I=\Omega_{I,\text{RWA}}$ and $\lambda=1/2\omega_d$ and the dashed line correspond to the truncated series solution for $k\leq 14$. (b) Corrected and uncorrected error as a function of time for a 20 ns gate. In (a) and (b) the carrier phase $\phi=0$. (c)-(f) Heatmaps of the same simulation as in (b) as a function of the carrier phase. (c) and (d) show the error for the uncorrected pulse parameters and algebraic result respectively. (e) and (f) show the results for the truncated series solution with $k\leq 14$ corresponding to the parameter choices $\beta=\pi$ and $t_0=0$ respectively. The dashed lines in (d) and (e) indicate the times for which $t=t_0+nt_c$.}
\end{figure}

In Fig. \ref{fig:figure12} we investigate the accuracy of the correction terms for the zeroth-order Magnus expansion for $\omega_{01}/2\pi=\omega_d/2\pi=80$ MHz. Specifically, we compute the error $E$ between the desired RWA time evolution and zeroth-order Magnus expanded non-RWA time evolution for varying pulse parameters. The error between two time evolutions $U$ and $V$ is defined as $E=1-F$ with $F$ the fidelity defined by \cite{fidelity1, fidelity2}:

\begin{equation}
    F\left(U,V\right) = \frac{1}{6}\left(2 + \Tr(UV^\dagger)\Tr(U^\dagger V)\right).
\end{equation}

\noindent In Fig. \ref{fig:figure12}(a) the error is plotted as a function of the gate duration for $\phi=t_0=\beta=0$. First, we naively use the RWA pulse parameters in the zeroth-order Magnus expanded non-RWA Hamiltonian which we refer to as the uncorrected pulse parameters, i.e. $\Omega_I=\Omega_{I,\text{RWA}}$ and $\lambda=0$. For these pulse parameters, the error increases exponentially as the gate duration shortens. Whenever the gate is commensurate, i.e. $t_g=nt_c$ for an integer $n$, the error for the uncorrected pulse parameters is zero since the oscillating non-RWA terms in the Hamiltonian integrate exactly to zero. For the corrected parameters, the algebraic solution to Eq. \eqref{eq:magnus0-requirement} (solid lines) and the truncated series solution (dashed lines) are plotted. For the truncated series solution, the pulse parameters are computed up to $15^\text{th}$ order in $t_c/t_g$, i.e. $k \leq 14$. For the algebraic parameters the error is negligible and independent of the gate duration. For the truncated series solution the error increases suddenly for $t_g<2t_c$, which is ascribed to the truncation error due to the aforementioned convergence rate of the infinite series in Eq. \eqref{eq:magnus0-requirement}. In Fig. \ref{fig:figure12}(b) we plot the error as a function of time for a 20 ns gate with the same parameters as in Fig. \ref{fig:figure12}(a) for the uncorrected and corrected pulse parameters. We see that, as detailed in \mbox{App. \ref{app:beta0}}, the corrected time evolution intersects the RWA evolution every Magnus period as well as at the end of the gate. 

In Figs. \ref{fig:figure12}(c)-(f) we perform the same simulations as in Fig. \ref{fig:figure12}(b) as a function of $\phi$ and $t_0$. In \mbox{Figs. \ref{fig:figure12}(c) and (d)} we compute the error for the uncorrected pulse parameters and the algebraic solution respectively. As expected, for the uncorrected pulse parameters the RWA and non-RWA evolutions do not intersect. For the algebraic pulse parameters the evolutions intersect every Magnus period and at $t=t_g$. The absolute time at which the evolutions intersect varies, since a change in carrier phase $\phi$ requires a shift in the start of the integration window $t_0$ as we require $\beta=2\omega_d t_0 + 2\phi=0$ for the algebraic solution. In \mbox{Fig. \ref{fig:figure12}(e)} we use the truncated series solution corresponding to $\beta=\pi$, where we observe that low errors are only achieved when $t_0 \approx 0$ corresponding to $\phi \approx \beta/2 = \pi/2$. For other values of the carrier phase, $t_0 \neq 0$ such that the RWA and non-RWA evolutions have already diverged before $t=t_0$ and will not intersect anymore. In Fig. \ref{fig:figure12}(f) we therefore set $t_0=0$ for each value of $\phi$ by varying $\beta$. As expected, the RWA and non-RWA time evolutions intersect at $t=N_ct_c$, but diverge in the uncorrected time interval $t\in[N_ct_c,t_g]$.


Even though we have now only considered the zeroth-order Magnus expansion, the results in Fig. \ref{fig:figure12} already provide us with a clear intuition for the ideal pulse parameters beyond the RWA. For each gate, the carrier phase increments with $w_dt_g$ with respect to the previous gate, such that each gate is represented by a different horizontal linecut in Fig. \ref{fig:figure12}. Therefore, we conclude that high-fidelity operations beyond the RWA are enabled through the existence of the exact algebraic solution shown in Fig \ref{fig:figure12}(d), as this is the only solution for which the pulse parameters are independent of the carrier phase and the error is low for each carrier phase.

\subsection{Higher Order Magnus Expansion \label{sec:theory-higher-order}}
While the zeroth-order Magnus term $\overline{H}^{(0)}$ is often the most significant contribution to the average Hamiltonian $\overline{H}$, it is not sufficient to accurately model the time evolution. In this section we therefore extend the analysis to the first-order Magnus approximation and full time evolution. Analytically calculating the ideal pulse parameters for the first-order Magnus approximation requires solving several challenges. Here, we lay-out the general steps to computing the pulse parameters in the first-order Magnus approximation, and full details can be found in \mbox{App. \ref{app:first-order-magnus-expansion}}. 

First, we need to solve the double integrals in the first-order Magnus expansion in Eq. \eqref{eq:magnus-definition-1}. We use similar methods as for the zeroth-order Magnus term to obtain infinite series expressions for these integrals, providing expressions for the drive strength $\Omega_I$, PPP $\lambda$ and detuning $\Delta$. These expressions form a non-linear system of equations, which is challenging to solve. Therefore, we use fixed-point iteration to iteratively compute the pulse parameters. Even though we found this approach to numerically converge well, we found the computed pulse parameters to result in large gate errors due to non-negligible terms in the uncorrected time intervals. To resolve this problem, we split the carrier signal into a commensurate and incommensurate term, and absorb the incommensurate term into the pulse envelopes while the commensurate term acts as the new carrier signal. This enables integrating over the full gate duration at the cost of slower convergence of the infinite series expressions for the pulse parameters. We then define the ideal pulse parameters as those that minimize the error averaged over the carrier phases $\phi$, which we approximate as the mean of the pulse parameters as a function of $\phi$.

\begin{figure}[t]
\centering
\includegraphics[width=\linewidth]{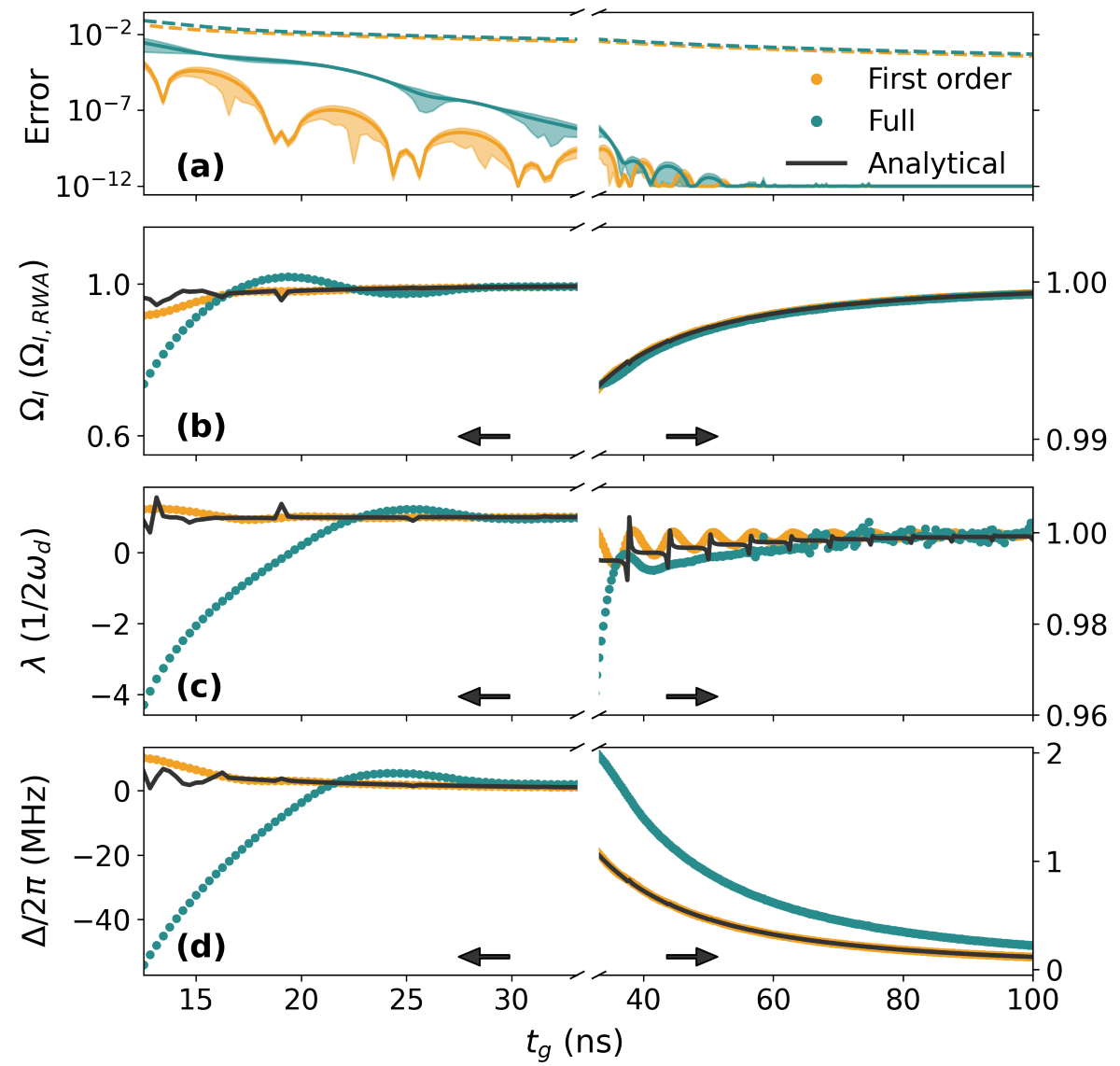}
\caption{\label{fig:figure3} (a)-(d) Optimized error, drive strength $\Omega_I$, PPP $\lambda$ and detuning $\Delta$ respectively for $\omega_{01}/2\pi = 80$ MHz. The cost function for the optimizer averages over 12 different carrier phases, i.e. $N=11$ in \mbox{Eq. \eqref{eq:cost_function}}. In (a), the solid lines indicate the mean error and the shaded area indicates the minimum and maximum error for the different carrier phases. The dashed lines indicate the errors for the uncorrected pulse parameters, i.e. $\mathcal{P}=\mathcal{P}_\text{RWA}$. The solid lines in (b)-(d) indicate the pulse parameters calculated using the approach detailed in App. \ref{app:first-order-magnus-expansion}. The PPP in (c) is calculated in units of $1/2\omega_d$, where $\omega_d$ is computed using the detunings shown in (d).}
\end{figure}

In Fig. \ref{fig:figure3} we numerically optimize the pulse parameters for an $X_\pi$ gate for the first-order Magnus approximation and for the full time evolution and compare them with the analytically computed ideal pulse parameters. The time evolution corresponding to the first-order Magnus approximation is calculated using Eqs. \eqref{eq:magnus-definition-1} and \eqref{eq:magnus-definition-2}, and the full time evolution is calculated using an ODE solver based on the LSODA algorithm \cite{lsoda, odepack, scipy}. The cost function of the optimizer corresponds to the error averaged over a range of different carrier phases $\phi$:

\begin{equation}
\label{eq:cost_function}
    C = \sum_{k=0}^{N} \Big(1 - F\big(U(\phi=\pi k/(N+1)), X_\pi\big) \Big),
\end{equation}

\noindent which ensures that the optimizer converges to pulse parameters that have a low error for all carrier phases $\phi$. The optimized pulse parameters for the first-order Magnus approximation match well with the analytically computed pulse parameters for gate durations longer than 5 Magnus periods, which corresponds to approximately \mbox{30 ns} for the qubit frequency used here. The drive strength and PPP of the full time evolution match well with the theoretical pulse parameters. However, the detuning is significantly higher, which we ascribe to the non-negligible influence of higher order terms in the Magnus expansion. For gate durations shorter than 5 Magnus periods the error increases exponentially for both the first-order Magnus approximation and the full time evolution. Additionally, the error in the full time evolution is considerably higher for these shorter gate durations, and the analytically computed pulse parameters are no longer accurate. Most likely, this occurs due to the increasing dependence of the time evolution on the carrier phase as a result from the increasing influence of higher order terms in the Magnus expansion. 

We conclude that, while the first-order Magnus approximation is insufficient to accurately model the time evolution, the correction terms effectively suppress the counter-rotating terms in the Hamiltonian for gate durations above 5 Magnus periods. For shorter gate durations, the dependence of the time evolution on the carrier phase becomes too significant, and as a result the gate error increases exponentially.

\subsection{Implementing a Universal Gate Set \label{sec:theory-universal-gate-set}}
Until now, we restricted our analysis to the implementation of $X_\pi$ gates. For a universal gate set, it suffices to have an $X_{\pi/2}$ gate and arbitrary $Z$ rotations \cite{book_williams, virtual_z_gates}. Here, we discuss a more general case in which we allow arbitrary rotation angles about arbitrary rotation axes in the $XY-$plane of the Bloch sphere in addition to virtual $Z$ gates \cite{virtual_z_gates}. The rotation axis $\theta$ is implemented by rotating the in-phase and quadrature signals according to \mbox{Eq. \eqref{eq:drive_term_definition}}. This modifies the drive term in the rotating frame according to:

\begin{equation}
\label{eq:theta_drive_rotation}
\begin{split}
    \mathcal{E}_I(t) &\mapsto \cos(\theta)\mathcal{E}_I(t) - \sin(\theta)\mathcal{E}_Q(t), \\
    \mathcal{E}_Q(t) &\mapsto \cos(\theta)\mathcal{E}_Q(t) + \sin(\theta)\mathcal{E}_I(t).
\end{split}
\end{equation}

\noindent Following Eqs. \eqref{eq:hamiltonian-terms} and \eqref{eq:fundamental-equation} and by making the transformation $\ket{1} \mapsto e^{-i\theta}\ket{1}$ it can be shown that the requirements that need to be satisfied to minimize the error remain the same, and that only $\beta$ changes according to $\beta \mapsto \beta - 2\theta$. Hence, the ideal pulse parameters are independent of the rotation axis. 

From the calculations in sections \ref{sec:theory-0th-order} and \ref{sec:theory-higher-order} and Apps. \ref{app:zeroth-order-magnus-expansion}-\ref{app:first-order-magnus-expansion} it is apparent that the ideal pulse parameters depend non-linearly on the ratio $\Omega_I/\Omega_{I,\text{RWA}}$ such that they also depend non-linearly on the desired rotation angle. Within the RWA, the ideal drive strength is a linear function of the desired rotation angle, meaning that the determination of the drive strength for a specific rotation angle immediately gives the ideal drive strength for any other rotation angle. Beyond the RWA this no longer holds, and the ideal pulse parameters are a non-trivial function of the desired rotation angle and must be determined for each rotation angle separately. 

\begin{figure}[t]
\centering
\includegraphics[width=\linewidth]{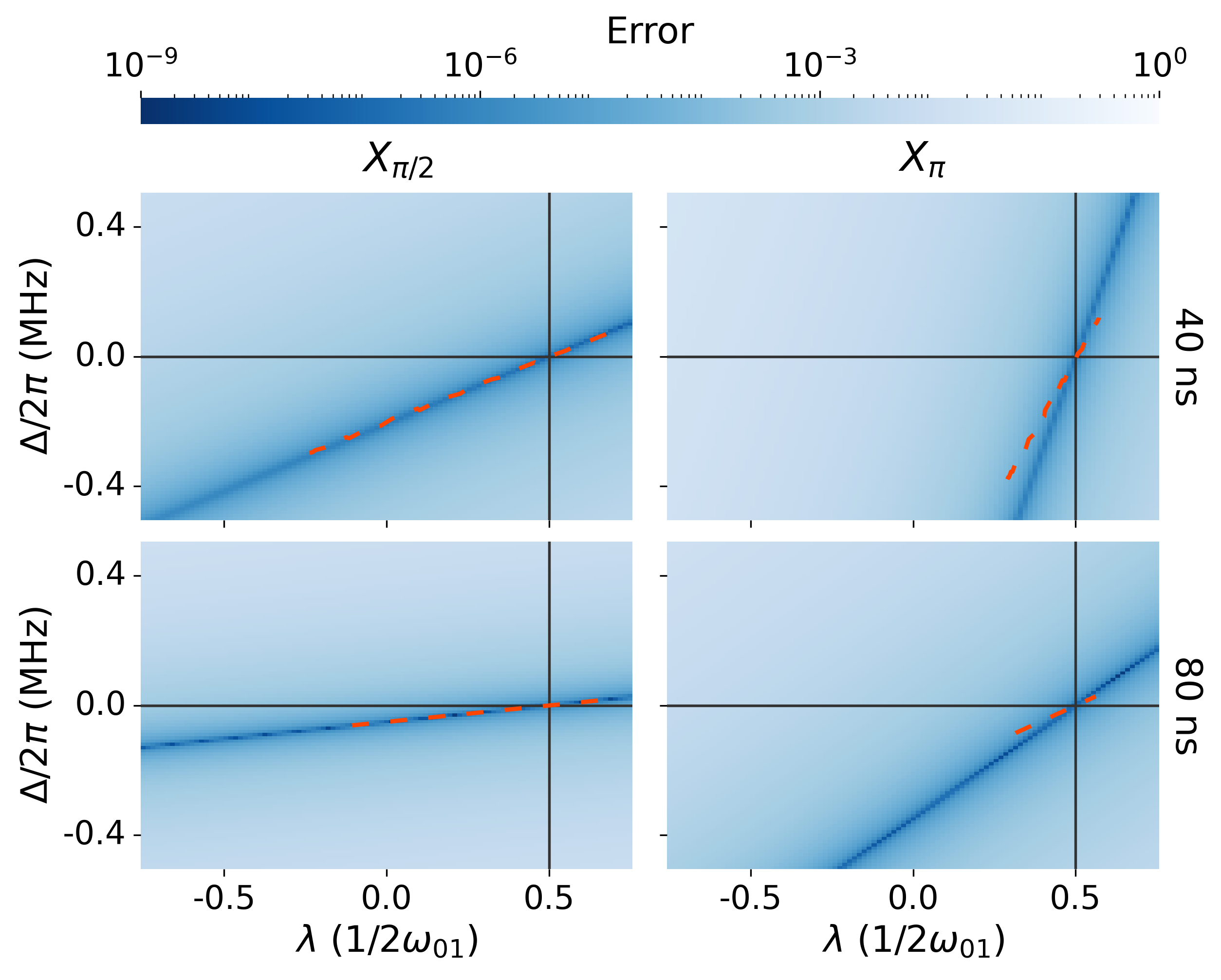}
\caption{\label{fig:figure6} Minimum error versus $\lambda$ and $\Delta$ for a two-level system with $\omega_{01}/2\pi = 80$ MHz. For each point, the error is minimized as a function of the drive strength $\Omega_I$ according to the cost function in Eq. \eqref{eq:cost_function} with $N=12$. The first (second) row shows the results for a 40 (80) ns gate, and the first (second) column shows the results for an $X_{\pi/2}$ ($X_\pi$) gate. The crossings at $\Delta=0$ and $\lambda=1/4\omega_{01}$ are indicated with black solid lines. The red dashed lines indicate the analytically calculated pulse parameters corresponding to $t_0=0$ and \mbox{varying $\beta$.}}
\end{figure}

To further emphasize the dependence of the pulse parameters on the desired rotation angle, we numerically compute the error as a function of $\lambda$ and $\Delta$ in \mbox{Fig. \ref{fig:figure6}}. For each point, we optimize the drive strength $\Omega_I$ according to the cost function in Eq. \eqref{eq:cost_function}, such that each point represents the minimum achievable error for a given $\lambda$ and $\Delta$. We perform these simulations for $X_{\pi/2}$ and $X_\pi$ gates and for $t_g = 40$ and $80$ ns of which the results are shown in \mbox{Fig. \ref{fig:figure6}}. Even though there exists a global minimum in this parameter space, we find that there also exists a large contour on which the error is low. The existence of this contour is explained by the freedom to choose $t_0$ and $\phi$ when computing the pulse parameters. The dashed lines in \mbox{Fig. \ref{fig:figure6}} represent the ideal pulse parameters calculated using $b_-=t_0=0$, $b_+=N_ct_c$ and along varying $\beta$. Even though this approach is not the most accurate for calculating the global minimum as done in Fig. \ref{fig:figure3}, these pulse parameters still yield errors below $10^{-6}$ since the uncorrected time interval $t \in [N_ct_c,t_g]$ is small relative to the total gate duration. The computed pulse parameters do not align exactly with the low-error contours, which we suspect to be a result from higher-order terms in the Magnus expansion. Furthermore, they only account for a portion of the low-error contour. Other parts of the contour are explained by satisfying Eq. \eqref{eq:fundamental-equation} over smaller time intervals, for instance $b_-=t_0=0$ and $b_+=(N_c-1)t_c$.

While these contours depend on the gate duration and drive strength, Fig. \ref{fig:figure6} suggests that they pass through the point $\Delta=0$, $\lambda=1/4\omega_{01}$ seemingly independently of the gate duration and desired rotation angle. From the theory in section \ref{sec:theory-higher-order} and App. \ref{app:first-order-magnus-expansion} we understand that, if such a crossing exists, it should happen at $\Delta=0$ as this indicates that the $\sigma_z$ terms in the first-order Magnus expansion integrate to zero, which only leaves the zeroth-order correction terms in which $\lambda$ is independent of the drive strength. For short gate durations, this crossing might occur at a different point in the parameter space due to the influence of higher-order terms in the Magnus expansion. Additionally, for a realistic system this crossing can also occur at a different point due to the influence of higher levels. Numerically, we find that the crossing at $\Delta=0$ and $\lambda=1/4\omega_{01}$ results in errors below $10^{-6}$ for gate durations as short as 2.8 Larmor periods, showing that for sufficiently long gate durations low errors can be achieved by assuming that $\Delta$ and $\lambda$ are independent of the drive strength and by finding the crossing of the low-error contours. This reduction in the dimensionality of the optimization problem eases the complexity of experimentally calibrating the pulse parameters.

\subsection{Strongly Anharmonic Multi-Level System \label{sec:theory-strongly-anharmonic-system}}
In this section, we derive and verify an additional correction term to the drive pulses arising from non-computational levels. We truncate the system at four levels under the assumption that the system is strongly anharmonic and the coupling to higher levels is sufficiently weak. For a fluxonium qubit the even transitions are parity forbidden at the operating point, such that we only have to consider drive terms for odd transitions. The drive Hamiltonian for such a system is a straightforward extension of Eq. \eqref{eq:two-level-hamiltonian} with the notable difference that we now allow the drive frequency $\omega_d(t)$ to be time-dependent. We expect that time-modulating the drive frequency is in general experimentally feasible due to our assumption that the qubit frequency is low. The Hamiltonian in the rotating frame $V(t)=\exp{i\sum_j j(\omega_d(t)t+\phi)\ket{j}\bra{j}}$ now becomes:

\begin{equation}
\label{eq:drive_hamiltonian}
\begin{split}
    &\tilde{H}/\hbar = \sum_{0<j<4} (\omega_{0j}-j\omega_d^\prime(t)) \ket{j}\bra{j} + \\
    &\sum_{jk\in \{01,12,23,03\}} A_x^{jk}(t)\eta_{jk} \sigma_x^{jk} + A_y^{jk}(t)\eta_{jk}\sigma_y^{jk}.
\end{split}
\end{equation}

\noindent Here, $\omega_d^\prime(t) = \omega_d(t) + t\dot{\omega}_d(t)$, $\omega_{0j}$ is the energy of the $j$-th level, and $\eta_{jk} = \abs{\bra{j}\hat{n}\ket{k}}/\abs{\bra{0}\hat{n}\ket{1}}$ is the relative drive strength of the transition $j \leftrightarrow k$ normalized by the drive strength of the qubit transition, i.e. $\eta_{01}=1$. In this work, we focus on charge driving such that the relative drive strengths are given by the charge matrix elements. Furthermore, $\sigma_x^{jk}=\ket{j}\bra{k}+\text{h.c.}$, $\sigma_y^{jk}=-i\ket{j}\bra{k}+\text{h.c.}$, $A_{x,y}^{01}(t)=A_{x,y}^{12}(t)=A_{x,y}^{23}(t)=A_{x,y}(t)$ and: 

\begin{equation}
\begin{split}
    A_x^{03}(t) &= \cos(2\omega_d(t)t+2\phi)A_x(t) - \\
    &\hspace{0.45cm} \sin(2\omega_d(t)t+2\phi)A_y(t), \\
    A_y^{03}(t) &= \cos(2\omega_d(t)t+2\phi)A_y(t) + \\
    & \hspace{0.45cm} \sin(2\omega_d(t)t+2\phi)A_x(t).
\end{split}
\end{equation}

\noindent The 03 drive term has a different time-complexity in this frame compared to the other drive terms because it does not act on neighboring levels. We perform an adiabatic transformation to derive a first-order effective two-level Hamiltonian, see App. \ref{app:fluxonium_as_tls} for more details. The effect of non-computational levels can be suppressed by setting a time-dependent detuning according to:

\begin{equation}
\label{eq:time_dependent_detuning}
    \Delta^\prime(t) \equiv \omega_{01} - \omega_d^\prime(t) = \frac{\mathcal{E}_I^2(t)+\mathcal{E}_Q^2(t)}{2}\left(\frac{\eta_{12}^2}{\alpha_2} - \frac{\eta_{03}^2}{\alpha_3}\right),
\end{equation}

\noindent with $\alpha_j = \omega_{0j} - j\omega_{01}$. The corresponding drive frequency $\omega_d(t)$ can be obtained by solving the differential equation in Eq. \eqref{eq:time_dependent_detuning} such that we obtain the detuning $\Delta(t)$ in the lab frame:

\begin{equation}
\label{eq:time_dependent_detuning2}
    \Delta(t) = \frac{1}{t}\int dt \Delta^\prime(t).
\end{equation}

Notice that, following Eq. \eqref{eq:theta_drive_rotation}, this correction term is independent of the desired rotation angle and axis. As detailed in App. \ref{app:fluxonium_as_tls}, the adiabaticity parameter of this transformation is $\max \{\eta_{12}\Omega_I/\alpha_2, \eta_{03}\Omega_I/\alpha_3 \}$. This implies that we strictly require the relative anharmonicity of the system to be much larger than the drive strength, i.e. $\alpha_2/\eta_{12}, \alpha_3/\eta_{03} \gg \Omega_I$. For the system studied here, $\Omega_I/2\pi \sim 100$ MHz, $\eta_{jk} \sim 1-20$ and $\alpha_j/2\pi \sim 1-5$ GHz. Hence, the adiabaticity parameter can approach the order of unity, which compromises the accuracy of the effective Hamiltonian. Therefore, we introduce an additional pulse parameter $\Omega_\Delta$ which rescales $\Delta(t)$ according to $\Delta(t) \mapsto \Omega_\Delta \Delta(t)$ or equivalently $\Delta^\prime(t) \mapsto \Omega_\Delta \Delta^\prime(t)$. Additionally, we allow a rescaling of the drive amplitude according to $\Omega_I \rightarrow \epsilon\Omega_I$.

\begin{figure}[t]
\centering
\includegraphics[width=\linewidth]{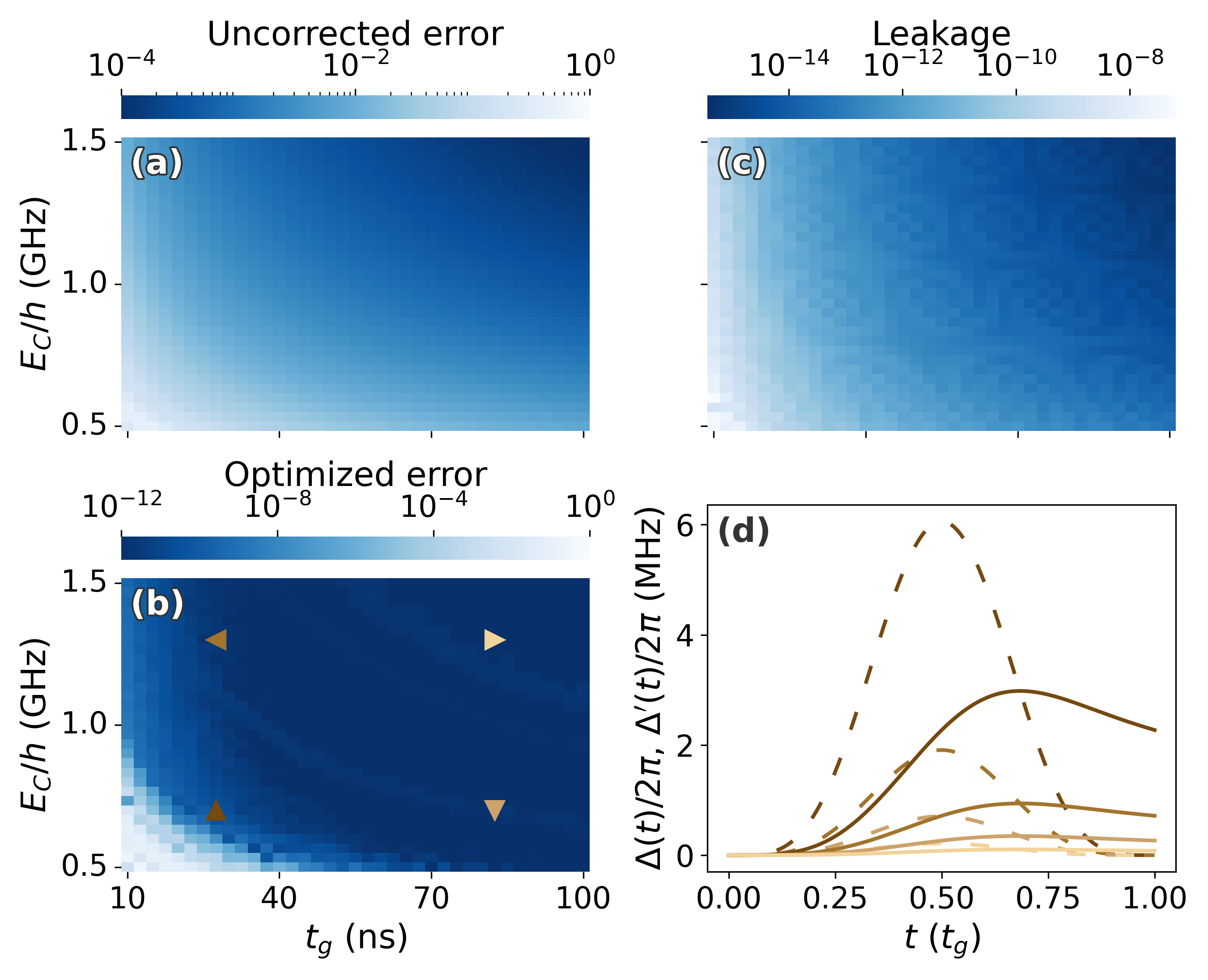}
\caption{\label{fig:figure5} Error between the time evolutions corresponding to a four-level and two-level system for a fluxonium qubit with $E_J/h =$ 5 GHz, $E_L/h =$ 1 GHz and as a function of $E_C$ and the gate duration $t_g$. The carrier phase is fixed to 0. (a) Uncorrected error, i.e. $\Omega_\Delta=0$ and $\epsilon=1$. (b) Optimized error as a function of $\Omega_\Delta$ and $\epsilon$. (c) Leakage to non-computational states for the optimized parameters calculated using Eq. \eqref{eq:leakage_definition}. (d) Optimized $\Delta(t)$ (solid lines) and $\Delta^\prime(t)$ (dashed lines) calculated using Eqs. \eqref{eq:time_dependent_detuning} and \eqref{eq:time_dependent_detuning2} respectively for four points in the parameter space as indicated in (b).}
\end{figure}

To verify this correction term, we numerically minimize the error between the time evolution operator in the computational subspace for the two-level and four-level Hamiltonians for a realistic fluxonium system as a function of $\Omega_\Delta$ and $\epsilon$. We fix $E_J/h =$ 5 GHz and $E_L/h =$ 1 GHz and sweep $E_C$ to make the fluxonium heavier or lighter. We also sweep the drive strength by varying the gate duration $t_g$. The fluxonium energy levels and charge matrix elements corresponding to these parameters can be found in App. \ref{app:fluxonium_as_tls} as well as the fitted pulse parameters $\Omega_\Delta$ and $\epsilon$. In Fig. \ref{fig:figure5}(a) and (b) we plot the uncorrected and corrected error respectively. For the uncorrected error we set $\Omega_\Delta=0$ and $\epsilon=1$. We find that not correcting for non-computational levels results in errors above $10^{-4}$ and that the correction term in Eq. \eqref{eq:time_dependent_detuning} can effectively suppress the influence from these levels for the majority of the parameter range. However, for short gates and heavy fluxonium parameters the error increases sharply. We attribute this to the higher-order terms in the adiabatic transformation. In Fig. \ref{fig:figure5}(c) we compute the leakage rate to non-computational levels for the optimized pulse parameters which we define as:

\begin{equation}
\label{eq:leakage_definition}
    \gamma_L(U) = \frac{1}{6}\sum_{\psi_f \in \{2,3\}}\sum_{\psi_i \in \{\pm x, \pm y, \pm z\}} \abs{\bra{\psi_f}U\ket{\psi_i}}^2.
\end{equation}

\noindent We see that leakage is negligible for this parameter range. In Fig. \ref{fig:figure5}(d) we illustrate the magnitude of the correction term by plotting the time-dependent detunings in the lab frame and rotating frame for four different parameters as indicated in Fig. \ref{fig:figure5}(c). 

We now have two corrections to the drive frequency $\omega_d(t)$: a time-independent correction to correct for the non-RWA terms as derived in section \ref{sec:theory-higher-order}, and a time-dependent correction for the non-computational levels as derived in this section. To distinguish these two contributions, we explicitly write the drive frequency as a sum of these two contributions:

\begin{equation}
    \omega_d(t) = \omega_{01} - \Delta - \Delta(t),
\end{equation}

\noindent where $\Delta$ is calculated in section \ref{sec:theory-higher-order} and $\Delta(t)$ is given in Eq. \eqref{eq:time_dependent_detuning2}. We highlight that, in the following section, the time-independent detuning $\Delta$ is calibrated directly, while the time-dependent detuning $\Delta(t)$ is calibrated through the parameter $\Omega_\Delta$.

Time-modulating the drive frequency comes at a cost, since the non-stationary terms in the two-level Hamiltonian also obtain a time-dependent drive frequency. In the derivation of the correction terms for the non-RWA terms in the Hamiltonian in sections \ref{sec:theory-0th-order} and \ref{sec:theory-higher-order} we assumed that the drive frequency is time-independent. Setting a time-dependent drive frequency therefore potentially compromises the accuracy of the ideal pulse parameters calculated in this work. This challenge may be resolved by absorbing the time-dependence of $\omega_d(t)$ into the pulse shapes, similar to the approach detailed in App. \ref{app:first-order-magnus-expansion}, but we leave this for future work. We are therefore forced to assume that the drive frequency varies sufficiently slow such that we can assume that $\lambda \propto 1/2\omega_d(t)$. To prevent having to time-modulate $\lambda$, we redefine the quadrature pulse envelope as:

\begin{equation}
    \mathcal{E}_Q(t) = \lambda \frac{\partial \mathcal{E}_I(t)}{\partial t} \frac{\omega_{01} - \Delta}{\omega_{01} - \Delta - \Delta(t)}.
\end{equation}

\section{Experimental Implementation \label{sec:experiment}}
Calibrating the four pulse parameters $\Omega_I$, $\lambda$, $\Delta$ and $\Omega_\Delta$ for each desired rotation angle in an actual experiment poses a major challenge, since the system of equations that needs to be solved to obtain the minimum error is coupled in the pulse parameters. As a result, it is not possible to calibrate the pulse parameters one by one, but instead they need to be calibrated simultaneously. This can be achieved experimentally by using an optimization algorithm such as optimized randomized benchmarking for immediate tune-up (ORBIT) \cite{orbit}, which uses a classical algorithm to minimize an experimentally obtained proxy for the gate error based on fixed length randomized benchmarking sequences. As these black-box minimization algorithms are typically resource-intensive and might not converge (to a global minimum), we make use of the findings in section \ref{sec:theory-universal-gate-set} to reduce the dimensionality of this optimization problem and derive deterministic calibration protocols for $\pi$ and $\pi/2$ rotations.

We use pseudo-identity circuits to amplify specific errors in the gates. A pseudo-identity circuit is a circuit that implements an identity when there are no errors in the operation. When there are errors, they are amplified by repeatedly applying the pseudo-identity circuit. For example, we use the pseudo-identity circuit $X_\pi - X_\pi$ to amplify over/under-rotations arising mostly from incorrectly calibrating $\Omega_I$. When repeating this circuit $M$ times, the measured signal will oscillate as a function of $M$, where the oscillation amplitude and frequency depend on the magnitude of the error. Specifically, we use the difference between the maximum and minimum measured signal as a metric for the magnitude of the error. We additionally use the circuit $X_{\pi/2} - X_{\pi/2} - X_{\pi/2} - X_{\pi/2}$ to calibrate the drive strength $\Omega_I$ for the $\pi/2$ gate. We use the circuits $X_\pi - Z - X_\pi - Z$ and $X_{\pi/2} - Z - X_{\pi/2} - Z$ to amplify phase errors \cite{pseudo_identity_circuit}. For all circuits, the qubit is prepared in the $\ket{0}$ state. For the under/over-rotation pseudo-identity circuits the qubit is measured in the $Y$-basis and for the phase error pseudo-identity circuits in the $X$-basis. 

We use four calibration protocols to study the minimum achievable error using the pulse shapes developed in this work as well as the accuracy of deterministic calibration protocols:

\begin{enumerate}[(P1)]
    \item \textit{Ideal RWA}: we assume that the RWA holds exactly and that the system is a true two-level system, and only calibrate the drive strength for the $X_\pi$ and $X_{\pi/2}$ gates while setting $\lambda=\Delta=\Omega_\Delta=0$.
    \item \textit{Non-RWA only}: we assume that there are only non-RWA errors, and make a heatmap of the phase error versus $\lambda$ and $\Delta$ for each desired rotation angle. We intersect the low-error contours to find the optimal values for $\lambda$ and $\Delta$. We calibrate the drive strength $\Omega_I$ before and after the characterization of the phase errors.
    \item \textit{Non-RWA + MLS (multi-level system)}: we correct for non-RWA errors by fixing $\Delta=0$ and $\lambda=1/4\omega_{d}$ and correct for non-computational levels by calibrating $\Omega_\Delta$ for the $X_\pi$ and $X_{\pi/2}$ gates individually using the phase error pseudo-identity circuits. We calibrate the drive strength $\Omega_I$ before and after calibrating $\Omega_\Delta$.
    \item \textit{ORBIT}: we perform ORBIT as a function of all the pulse parameters to obtain the minimum achievable error.
\end{enumerate}

\noindent We experimentally implement these protocols on a \mbox{98.97 MHz} fluxonium qubit with average coherence times of $T_1 = 75$ $\mu$s and $T_{2\text{E}} = 37$ $\mu$s. All experiments are averaged over the carrier phase by incrementing the carrier phase by 1 degree in each repetition of the circuit. The protocols are implemented with gate durations $t_g$ of 13.3 ns, 20 ns, 26.7 ns, 33.3 ns and 40 ns. If the qubit is driven on resonance, the corresponding number of Magnus periods $N_c$ are 2.64, 3.96, 5.28, 6.60 and 7.92. More details about the experimental setup can be found in App. \ref{app:device_and_setup}. 

\begin{figure}[b]
\centering
\includegraphics[width=\linewidth]{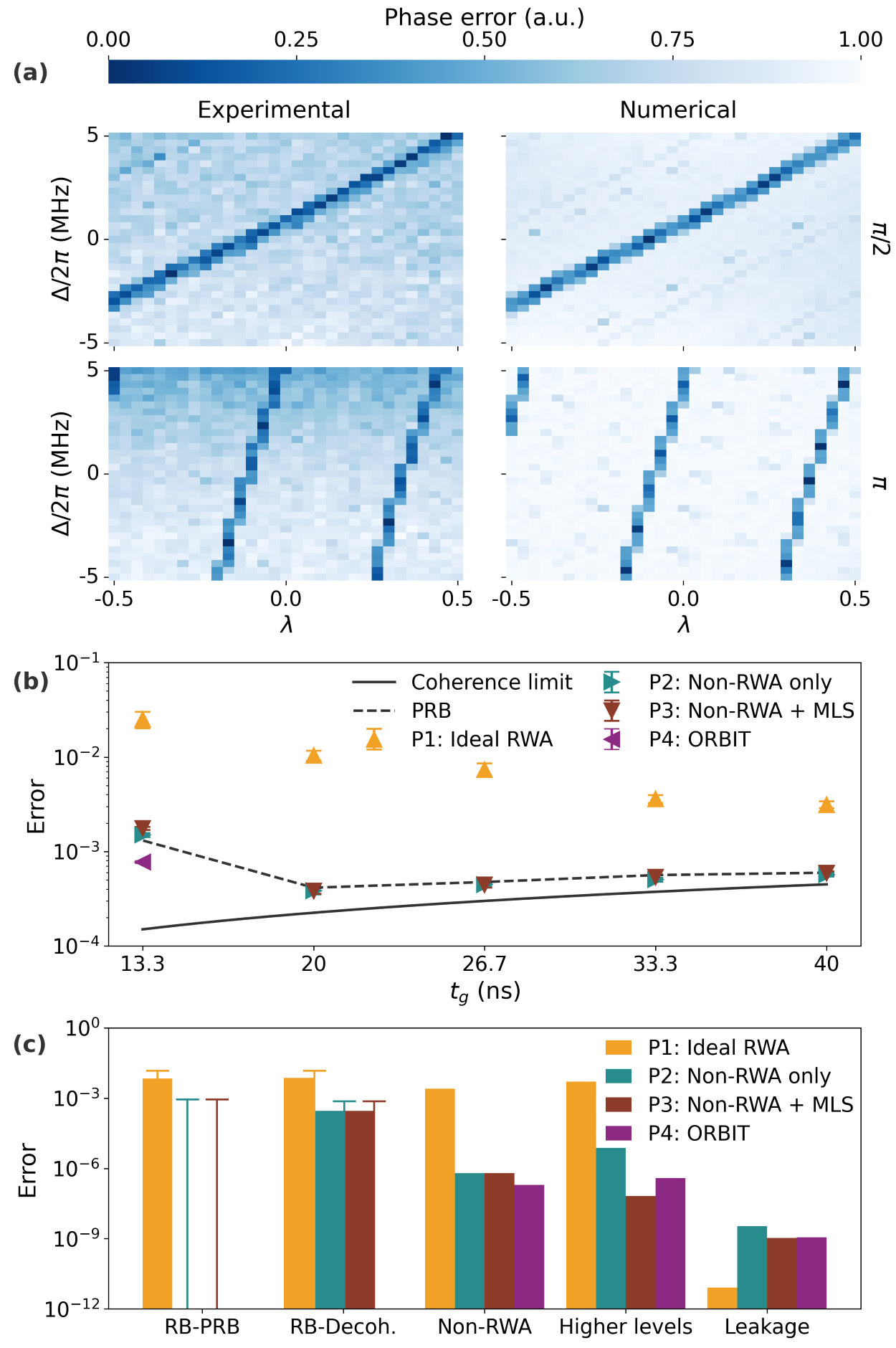}
\caption{\label{fig:experimental-results} (a) Heatmaps of the phase error versus $\lambda$ and $\Delta$ for $t_g=26.7$ ns. The first (second) row contains the results for the $\pi/2$ ($\pi$) gate and the first (second) column contains the experimental (numerical) results. (b) Measured average gate fidelity using RB for all gate durations and protocols. The solid line indicates the coherence limit, and the dashed line indicates the measured average incoherent error for protocol P3. ORBIT has only been run for $t_g=13.3$ ns. (c) Error budget for $t_g=26.7$ ns. All errorbars indicate one standard deviation.}
\end{figure}

In Fig. \ref{fig:experimental-results}(a) we plot the phase error heatmap for protocol P2 for $t_g=26.7$ ns. The heatmaps for the remaining gate durations can be found in App. \ref{app:extended_data}. We also compute these heatmaps numerically and find excellent agreement with the experimental data for all gate durations. In these heatmaps we noticed that the low-error contours still intersect at $\Delta=0$, but at lower values of $\lambda$ compared to the simulations in Fig. \ref{fig:figure6}. Therefore, we optimize $\lambda$ by performing a one-dimensional sweep of the phase error versus $\lambda$ with $\Delta=0$. Additionally, more than one low-error contour is visible in these heatmaps. Numerically, we find that only one of those contours (the ones that intersect at $\Delta=0$) correspond to low gate errors, and the other contours arise when the phase errors are so large that they drive a full $2\pi$ rotation such that the pseudo-identity circuit implements an identity.

To characterize the fidelity of the operations, we perform randomized benchmarking (RB) \cite{original_randomized_benchmarking, formal_randomized_benchmarking, formal_randomized_benchmarking2} and purity randomized benchmarking (PRB) \cite{purity_rb} with $2^{13}$ repetitions following all calibration protocols, of which the results are shown in Fig. \ref{fig:experimental-results}(b). RB and PRB provide the average gate fidelity and average incoherent gate fidelity respectively, and we are most interested in the difference between these quantities as it is a measure for the magnitude of the coherent errors. We also plot the coherence limit defined by \cite{fidelity_decoherence, heinsoo_thesis}:

\begin{equation}
    F_\text{decoh} = \frac{1}{2} + \frac{1}{6}\exp(-\frac{t_g}{T_1}) + \frac{1}{3}\exp(-\frac{t_g}{T_{2\text{E}}}).
\end{equation}

\noindent From Fig. \ref{fig:experimental-results}(b) it is apparent that the errors obtained using protocol P1 are not coherence limited, indicating that there are indeed large non-RWA errors and/or errors from non-computational levels that should be corrected. By comparing the RB and PRB results we see that the coherent errors for protocol P2 and P3 are negligible for $t_g \geq 20$ ns. However, we also notice that the errors are not coherence limited. We ascribe this to heating from the microwave pulses, as discussed in \mbox{App. \ref{app:extended_data}}. For $t_g=13.3$ ns none of the protocols achieve coherence limited fidelities. Furthermore, we see that the incoherent error is much larger than the coherence limit for this gate duration. We attribute this to an increase in the variance of the sequence fidelities of random Clifford circuits arising from the increase in the dependence of the time evolution on the carrier phase, see also \mbox{App. \ref{app:extended_data}}. We only run ORBIT for $t_g=13.3$ ns, as protocol P2 and P3 already achieve negligible coherent errors for all the other gate durations. While ORBIT improved the error compared to the deterministic calibration protocols, significant errors remain in the gate. Due to the unreliable measurement of the incoherent error for this gate duration it is not possible to conclude whether these remaining errors are coherent or incoherent.

For each gate duration and protocol we numerically calculate error budgets in order to estimate the minimum achievable error for each protocol and to determine whether the error is limited by non-RWA errors or non-computational levels. To calculate these error budgets, we simulate each calibration protocol for a two-level system and for a four-level system. The resulting error for the two-level system is the error ascribed to non-RWA errors, and the difference between the errors for the four-level system and two-level system to non-computational levels. Additionally, we calculate the leakage error for the four-level system according to Eq. \eqref{eq:leakage_definition}. We do not simulate protocol P3 for a two-level system as there are no higher levels to correct for, and we simply reuse the non-RWA error from protocol P2. 

The error budgets for $t_g=26.7$ ns are plotted in \mbox{Fig. \ref{fig:experimental-results}(c)} and the error budgets for the remaining gate durations can be found in App. \ref{app:extended_data}. The first data group shows the measured coherent error computed by subtracting the experimentally measured average incoherent gate fidelity from the experimentally measured average gate fidelity. Especially for the lowest gate durations this estimate can be inaccurate due to the aforementioned increased dependence of the time evolution on the carrier phase. We therefore plot a second estimate of the coherent error in the second data group by computing the difference between the average gate fidelity and the numerically calculated decoherence error $E_\text{decoh}=1-F_\text{decoh}$. The accuracy of this second estimate of the coherent error is limited due to the heating effects. The remaining three data groups contain the numerically computed estimates of the error rates. The total error rate estimated numerically is the sum of the three error rates. Even though we do not perform ORBIT experimentally for $t_g \geq 20$ ns we still simulate it for all gate durations to show the minimum achievable error using the pulse shapes developed in this work. 

From these error budgets it is apparent that the error for protocol P1 consists of an almost equal combination of non-RWA errors and errors from non-computational levels, and that the total simulated coherent error matches well with the measured coherent error. By using protocol P2 the non-RWA errors are suppressed by one to five orders of magnitude depending on the gate duration. The remaining error in protocol P2 is limited by errors from higher levels, even though these errors, somewhat surprisingly, are also residually suppressed when compared to protocol P1. For $t_g>20$ ns, protocol P3 can further suppress the errors from higher levels by several orders of magnitude. For shorter gate durations we found protocol P3 to be ineffective at improving the total error. Leakage is shown to be negligible for all parameters used here. The error budgets further show that ORBIT improves the total error by approximately 1.5 and 1 order of magnitude compared to protocol P2 and P3 respectively, see also App. \ref{app:extended_data}. This indicates that the deterministic calibration protocols are very effective at achieving errors that are close to the minimum achievable error using the pulse shapes developed in this work. Finally, the error budgets indicate that, using these deterministic calibration protocols consisting of relatively straightforward experiments, it is possible to achieve coherent error rates averaged over $\pi$ and $\pi/2$ gates below $10^{-6}$ for $t_g \geq 26.7$ ns, which corresponds to 2.64 Larmor periods. We note that, as can be seen in App. \ref{app:extended_data}, these errors are strongly limited by the $\pi$ gates, and we find that the error for the $\pi/2$ gates is 1-4 orders of magnitude smaller depending on the gate duration and calibration protocol. This is a result from the exponential increase in the error as the RWA becomes more invalid, as also shown in Fig. \ref{fig:figure3}.

\section{Outlook \label{sec:conclusion}}
In this work, we have derived and verified solutions to two challenges that arise when applying single-qubit operations in a regime where the RWA does not hold. Firstly, the time evolution, which is now challenging to compute since the Hamiltonian is no longer self-commuting, is calculated using the Magnus expansion. We further Taylor expand the pulse envelopes and restrict the integration windows to Magnus intervals to derive expressions for the ideal pulse parameters. The second challenge is that the time evolution now depends on the carrier phase such that, in general, each operation in a sequence of gates is different. We have shown that, in the zeroth-order Magnus approximation, there exist pulse parameters that correct exactly for the counter-rotating terms over the full gate duration independently of the carrier phase. This important and somewhat surprising result enables the implementation of high-fidelity operations beyond the RWA. We further derived the ideal pulse parameters in the first-order Magnus approximation. We found that the analytically calculated ideal pulse parameters match well with the numerically optimized pulse parameters for the first-order Magnus approximation. However, there are non-negligible effects from higher-order terms that compromise the accuracy of the calculated ideal pulse parameters for the full time evolution. We found that the minimum achievable error for an $X_\pi$ gate implemented on a two-level system using the pulse shapes derived in this work is negligible when the gate consists of more than five Magnus periods and increases exponentially for shorter gate durations. Achieving lower errors for gates faster than 5 Magnus periods may be achieved by restricting to commensurate gates or by implementing circularly polarized driving as discussed in Ref. \cite{circular_polarized_drive}.

While the ideal pulse parameters derived in this work are independent of the desired rotation axis, they depend strongly on the desired rotation angle. We found that a reduction in the dimensionality of the resulting optimization problem is possible due to the freedom to choose the start of the integration windows $t_0$ and carrier phase $\phi$. This insight was fundamental for the formulation of deterministic calibration protocols in the experimental part of this work.

We derived and verified an additional correction term arising from non-computational levels in a strongly anharmonic four-level system, inspired by the fluxonium superconducting qubit. We found that the influence from non-computational levels can be compensated by time-modulating the drive frequency. Even though the adiabaticity of the transformation that was used to derive this correction term is compromised, we found that this correction term is very effective at correcting for the influence from non-computational levels for sufficiently high gate durations and light fluxonium parameters. Time-modulating the drive frequency technically invalidates the optimal pulse parameters derived for the two-level system, since the derivations assume that the drive frequency is time-independent. This challenge may be resolved by absorbing the time-dependence of the drive frequency into the definition of the pulse envelopes, but this is left for future work.

Finally, we experimentally implemented the pulse shapes derived in this work on a 98.97 MHz fluxonium superconducting qubit. Notably, the implementation of these pulse shapes does not require any additional hardware complexities. Numerically, we found that the pulse shapes derived in this work can achieve coherent error rates below $10^{-6}$ averaged over $\pi$ and $\pi/2$ gates for a gate duration of 2.64 Larmor periods. Experimentally, we achieved a minimum error rate of $3.8 \cdot 10^{-4}$ for a gate duration corresponding to 1.98 Larmor periods. The error was limited by decoherence and heating effects from the microwave pulses. The heating effects could be mitigated by designing a less heavy fluxonium, by improving the filtering of the drive lines or by switching to flux driving. Switching to flux driving may additionally reduce the error from non-computational levels, as the fluxonium qubit levels couple much less strongly to non-computational levels through flux. Higher coherence may be achieved through improved design and fabrication of the device \cite{millisecond_coherence, mit_cz, high_coherence_fluxonium}. In future work, it would be insightful to study the experimental limits of the pulse shapes and calibration protocols developed in this work on an improved device and setup.

\section*{Acknowledgments}
The authors acknowledge the use of computational resources of the DelftBlue supercomputer, provided by Delft High Performance Computing Centre \cite{DHPC2024}. The authors further acknowledge Siyu Wang for valuable contributions to the design of the device.

The authors acknowledge support from the Dutch Research Council (NWO). Additionally, E.Y.H. acknowledges support from Holland High Tech (TKI), J.H. acknowledges support from NWO Open Competition Science M and T.V.S. acknowledges support from the Engineering and Physical Sciences Research Council (EPSRC) under EP/SO23607/1.

M.F.S.Z. performed the theoretical and numerical calculations. F.Y. designed the device. S.S., F.Y. and P.K. fabricated the device. M.F.S.Z., E.Y.H. and J.H. performed the experiments, and M.F.S.Z. analyzed the data. E.Y.H., T.V.S. and M.F.S.Z. developed and optimized the experimental setup. C.K.A. supervised the work. M.F.S.Z. wrote the manuscript with input from all authors.

\section*{Data Availability}
All numerical and experimental data is available through \cite{data_repository} and the code used for the numerical simulations and data processing is available through \cite{github_repo}.

\appendix

\onecolumngrid
\section{Zeroth-order Magnus expansion}
\label{app:zeroth-order-magnus-expansion}
In this section, we provide a detailed derivation of the pulse shape $s_Q(t)$. The proof is inspired by the Magnus-Taylor expansion introduced in Ref. \cite{exact_rwa}. Here, we only evaluate requirement (ii) of Eq. \eqref{eq:magnus0-three-requirements} since, as we will find, the pulse shape $s_Q(t)$ does not depend on the absolute drive strength $\Omega_I$. To satisfy requirement (ii), we need to solve integrals of the form:

\begin{equation}
\label{eq:taylor-integral-definitions}
\begin{split}
    \int_{b_-}^{b_+} dt_1 T\left[\mathcal{E}_i(t_1),t\right]\cos(2\omega_dt_1 + 2\phi) &= \sum_{k=0}^\infty \frac{\partial^k \mathcal{E}_i(t)}{\partial t^k} \int_{b_-}^{b_+} dt_1 \frac{(t_1-t)^k}{k!}\cos(2\omega_dt_1 + 2\phi), \\
    \int_{b_-}^{b_+} dt_1T\left[\mathcal{E}_i(t_1),t\right]\sin(2\omega_dt_1 + 2\phi) &= \sum_{k=0}^\infty \frac{\partial^k \mathcal{E}_i(t)}{\partial t^k} \int_{b_-}^{b_+} dt_1 \frac{(t_1-t)^k}{k!}\sin(2\omega_dt_1 + 2\phi).
\end{split}
\end{equation}

\noindent Here, $T\left[\mathcal{E}_i(t_1),t\right]$ denotes the Taylor expansion of $\mathcal{E}_i$ at time $t_1$ around time $t$, defined as:

\begin{equation}
    T\left[\mathcal{E}_i(t_1),t \right] = \sum_{k=0}^\infty \frac{\partial^k \mathcal{E}_i(t)}{\partial t^k} \frac{(t_1-t)^k}{k!}.
\end{equation}

\noindent The integrals in Eq. \eqref{eq:taylor-integral-definitions} can be computed by using integration by parts, and the solutions are:

\begin{equation}
\label{eq:taylor-integral-solutions}
\begin{split}
    &\int_{b_-}^{b_+} dt_1 \frac{(t_1-t)^k}{k!}\cos(2\omega_dt_1 + 2\phi) = \sum_{k'=k}^0 \left(\frac{1}{2\omega_d}\right)^{k-k'+1} \gamma_1(k-k',\beta) \left[\frac{(t_1-t)^{k'}}{k'!}\right]_{t_1=b_-}^{b_+} \equiv \mathcal{I_C}(k), \\
    &\int_{b_-}^{b_+} dt_1 \frac{(t_1-t)^k}{k!}\sin(2\omega_dt_1 + 2\phi) = -\sum_{k'=k}^0 \left(\frac{1}{2\omega_d}\right)^{k-k'+1} \gamma_2(k-k',\beta) \left[\frac{(t_1-t)^{k'}}{k'!}\right]_{t_1=b_-}^{b_+} \equiv \mathcal{I_S}(k),
\end{split}
\end{equation}

\noindent with $\gamma_{1,2}(k,\beta)$ defined in Eq. \eqref{eq:gamma-pm-def}. Substituting Eqs. \eqref{eq:taylor-integral-definitions} and \eqref{eq:taylor-integral-solutions} in requirement (ii) in Eq. \eqref{eq:magnus0-three-requirements} gives:

\begin{equation}
\label{eq:taylor-integral-solution-lambda}
\begin{split}
    \sum_{k=0}^\infty\frac{\partial^k \mathcal{E}_Q(t)}{\partial t^k}\left(\left[\frac{(t_1-t)^{k+1}}{(k+1)!}\right]_{t_1=b_-}^{b_+} - \mathcal{I_C}(k)\right) + \sum_{k=0}^\infty\frac{\partial^k \mathcal{E}_I(t)}{\partial t^k}\mathcal{I_S}(k) = 0.
\end{split}
\end{equation}

\noindent We note that $[(t_1-t)^{k'}/k'!]_{t_1=b_-}^{b_+}=0$ for $k'=0$, that $\gamma_1(k+1,\beta)=\gamma_2(k,\beta)$ and that:

\begin{equation}
    \mathcal{I_S}(k+1) = -\frac{1}{2\omega_d}\mathcal{I_C}(k) - \frac{1}{2\omega_d}\gamma_1(1,\beta)\left[\frac{(t_1-t)^{k+1}}{(k+1)!}\right]_{t_1=b_-}^{b_+}.
\end{equation}

\noindent With this we can rewrite Eq. \eqref{eq:taylor-integral-solution-lambda} as:

\begin{equation}
    \sum_{k=0}^\infty \Bigg\{ \left[\frac{(t_1-t)^{k+1}}{(k+1)!}\right]_{t_1=b_-}^{b_+}\left(\frac{\partial^k \mathcal{E}_Q(t)}{\partial t^k} - \frac{\gamma_1(1,\beta)}{2\omega_d}\frac{\partial^{k+1}\mathcal{E}_I(t)}{\partial t^{k+1}}\right) - \mathcal{I_C}(k)\left(\frac{\partial^k \mathcal{E}_Q(t)}{\partial t^k} + \frac{1}{2\omega_d}\frac{\partial^{k+1}\mathcal{E}_I(t)}{\partial t^{k+1}}\right) \Bigg\} = 0.
\end{equation}

\noindent From here, it is natural to set $\mathcal{E}_Q(t)=\lambda\frac{\partial \mathcal{E}_I(t)}{\partial t}$, since this allows for the straightforward derivation of a closed-form expression for $\lambda$:

\begin{equation}
\label{eq:taylor-integral-solution-lambda-final}
\begin{split}
    &\lambda \Big(\mathcal{E}_I(b_+)-\mathcal{E}_I(b_-)\Big) - \frac{\gamma_2(0,\beta)}{2\omega_d}\left[\frac{\partial^0 \mathcal{E}_I(t_1)}{\partial t_1^0}\right]_{t_1=b_-}^{b_+} - \sum_{k=0}^\infty \left(\frac{1}{2\omega_d}\right)^{k+1}\gamma_1(k,\beta)\left(\frac{1}{2\omega_d}+\lambda\right)\left[\frac{\partial^{k+1} \mathcal{E}_I(t)}{\partial t^{k+1}}\right]_{t_1=b_-}^{b_+} \\
    =& \lambda \Big(\mathcal{E}_I(b_+)-\mathcal{E}_I(b_-)\Big) - \sum_{k=0}^\infty \left(\frac{1}{2\omega_d}\right)^{k+1} \Bigg\{\lambda\gamma_1(k,\beta)\left[\frac{\partial^{k+1} \mathcal{E}_I(t)}{\partial t^{k+1}}\right]_{t_1=b_-}^{b_+} + \gamma_2(k,\beta)\left[\frac{\partial^{k} \mathcal{E}_I(t)}{\partial t^{k}}\right]_{t_1=b_-}^{b_+}\Bigg\} \\
    =&0.
\end{split}
\end{equation}

\noindent Notice that Eq. \eqref{eq:taylor-integral-solution-lambda-final} is analogous to the result obtained in Eq. \eqref{eq:magnus0-lambda-requirement} in the main text.

\onecolumngrid
\section{Special properties of $\beta=0$ and symmetric integration windows}
\label{app:beta0}
In this section, we derive the analytical solutions to Eq. \eqref{eq:fundamental-equation} corresponding to $\beta=0$ and symmetric integration windows around $t=t_g/2$.  We start by deriving these solutions for $\beta=0$ and we consider a slightly more general case in which gates start at an arbitrary time $\tau_0$, such that the gate is applied from $t=\tau_0$ to $t=\tau_0+t_g$. $t_0$ now represents the time relative to the start time of the gate, such that $\beta=2\omega_d(\tau_0+t_0)+2\phi$. First, we show that requirement (ii) in Eq. \eqref{eq:magnus0-three-requirements} holds for an arbitrary time interval $[t_0,t_0+t']$. For $\beta=0$, $\gamma_1(k,\beta)$ and  $\gamma_2(k,\beta)$ are only non-zero for odd and even values of $k$ respectively, such that we can rewrite Eq. \eqref{eq:magnus0-lambda-requirement} as:

\begin{equation}
\label{eq:beta0_lambda_req}
\begin{split}
    &\lambda \Big(s_I(t_0+t') - s_I(t_0)\Big) = \\
    &\sum_{k=0}^\infty \Bigg\{(-1)^{k} \lambda \left(\frac{1}{2\omega_d}\right)^{2k+2} \left[\frac{\partial^{2k+2}s_I(t_1)}{\partial t_1^{2k+2}} \right]_{t_1=t_0}^{t_0+t'} + (-1)^k \left(\frac{1}{2\omega_d}\right)^{2k+1}\left[\frac{\partial^{2k}s_I(t_1)}{\partial t_1^{2k}} \right]_{t_1=t_0}^{t_0+t'} \Bigg\}.
\end{split}
\end{equation}

\noindent We see that, if we set $\lambda = 1/2\omega_d$, the first term in the sum at \mbox{$k$-th} order is canceled by the second term in the sum at $k+1$-th order. The only remaining term is the second term in the sum at $k=0$, which cancels against the term on the first line. Hence, for $\beta=0$ and $\lambda=1/2\omega_d$ requirement (ii) holds for an arbitrary interval $[t_0,t_0+t']$ in the zeroth-order Magnus approximation. For requirement (iii) in Eq. \eqref{eq:magnus0-three-requirements} we can make a similar argument. Here, we will show that, on all relevant intervals:

\begin{equation}
\label{eq:beta0-three-requirements}
\begin{split}
    &(\text{i}) \; \int_{\tau_0}^{\tau_0+t_0}dt_1 \Big(\mathcal{E}_I(t_1)\cos(2\omega_dt_1+2\phi) + \mathcal{E}_Q(t_1)\sin(\omega_dt_1+2\phi)\Big) = 0, \\
    &(\text{ii}) \; \int_{\tau_0+t_0+N_ct_c}^{\tau_0+t_g}dt_1 \Big(\mathcal{E}_I(t_1)\cos(2\omega_dt_1+2\phi) + \mathcal{E}_Q(t_1)\sin(\omega_dt_1+2\phi)\Big) = 0, \\
    &(\text{iii}) \; \int_{\tau_0+t_0+(n-1)t_c}^{\tau_0+t_0+nt_c}dt_1 \Big(\mathcal{E}_I(t_1)\cos(2\omega_dt_1+2\phi) + \mathcal{E}_Q(t_1)\sin(\omega_dt_1+2\phi)\Big) = 0, \quad \forall \; n,
\end{split}
\end{equation}

\noindent such that, not only on the previously uncorrected intervals (i) and (ii) we require $\Omega_I=\Omega_{I,\text{RWA}}$, but also on the Magnus intervals that we could correct already (iii). We first simplify the integrals with general bounds $[t_0,t_0+t']$: 

\begin{equation}
\label{eq:beta0_omega_req}
\begin{split}
    &\int_{t_0}^{t_0+t'}dt_1 \Big(\mathcal{E}_I(t_1)\cos(2\omega_dt_1+2\phi) + \mathcal{E}_Q(t_1)\sin(\omega_dt_1+2\phi)\Big) \\
    =& \sum_{k=0}^\infty \left(\frac{1}{2\omega_d}\right)^{k+1} \Bigg\{  \left[\gamma_1(k,2\omega_dt_1+2\phi)\frac{\partial^{k}s_I(t_1)}{\partial t_1^{k}}\right]_{t_1=t_0}^{t_0+t'} - \lambda \left[\gamma_2(k,2\omega_dt_1+2\phi)\frac{\partial^{k+1}s_I(t_1)}{\partial t_1^{k+1}}\right]_{t_1=t_0}^{t_0+t'} \Bigg\} \\
    =&\frac{1}{2\omega_d}\Big(\mathcal{E}_I(t_0+t')\gamma_1(0,2\omega_d(t_0+t') + 2\phi) - \mathcal{E}_I(t_0)\gamma_1(0,2\omega_dt_0 + 2\phi)\Big) \stackrel{?}{=} 0.
\end{split}
\end{equation}

\noindent To go from the second to the third line in Eq. \eqref{eq:beta0_omega_req} we made use of the fact that $\gamma_1(k+1,\beta)=\gamma_2(k,\beta)$ such that each second term in the sum at \mbox{$k$-th} order cancels with the first term in the sum at $k+1$-th order if $\lambda=1/2\omega_d$ independently of the integration bounds. Requirement (iii) in Eq. \eqref{eq:beta0-three-requirements} is straightforwardly satisfied given $\beta=0$. Substituting the bounds for requirements (i) and (ii) in Eq. \eqref{eq:beta0_omega_req} gives:

\begin{equation}
\begin{split}
    &(\text{i})\; \Big(\mathcal{E}_I(\tau_0+t_0)\sin(2\omega_d(\tau_0+t_0) + 2\phi) - \mathcal{E}_I(\tau_0)\sin(2\omega_d\tau_0 + 2\phi)\Big) = 0, \\
    &(\text{ii})\; \Big(\mathcal{E}_I(\tau_0+t_g)\sin(2\omega_d(\tau_0+t_g) + 2\phi) - \mathcal{E}_I(\tau_0+t_0+N_ct_c)\sin(2\omega_d(\tau_0+t_0+N_ct_c) + 2\phi)\Big) = 0.
\end{split}
\end{equation}

\noindent Since $\sin(2\omega_d(\tau_0+t_0))=\sin(2\omega_d(\tau_0+t_0+N_ct_c) + 2\phi)=\sin(\beta)=0$, requirements (i) and (ii) in Eq. \eqref{eq:beta0-three-requirements} are satisfied if $\mathcal{E}_I(\tau_0)=\mathcal{E}_I(\tau_0+t_g)=0$, i.e. the pulse envelope should be zero at the start and the end of the gate. This requirement is satisfied for the cosine pulse envelopes in this work, and is generally satisfied for most pulse envelopes. 

The prove for the analytical solutions to Eq. \eqref{eq:fundamental-equation} in the case of symmetric integration windows follows similar steps, and relies on the assumption that $\frac{\partial^k \mathcal{E}_I(t)}{\partial t^k}$ is symmetric (anti-symmetric) around $t=t_g/2$ for even (odd) $k$. Using this symmetry property, $\gamma_1(k+1,\beta)=\gamma_2(k,\beta)$ and $\gamma_1(k+2,\beta)=-\gamma_1(k,\beta)$ we can rewrite Eq. \eqref{eq:magnus0-lambda-requirement} as:

\begin{equation}
\label{eq:symmetric_integration_windows_lambda}
    0 = \sum_{k=0}^\infty \left\{ \left(\frac{1}{2\omega_d}\right)^{2k+1}\lambda\gamma_1(2k,\beta)\left[\frac{\partial^{2k+1}s_I(t_1)}{\partial t_1^{2k+1}}\right]_{t_1=b_-}^{b_+} -  \left(\frac{1}{2\omega_d}\right)^{2k+2}\gamma_1(2k,\beta)\left[\frac{\partial^{2k+1}s_I(t_1)}{\partial t_1^{2k+1}}\right]_{t_1=b_-}^{b_+}\right\}.
\end{equation}

\noindent From which it is immediately clear that all terms cancel if $\lambda=1/2\omega_d$. The symmetric (anti-symmetric) terms in the remaining time intervals $t \in [\tau_0, b_-]$ and $t \in [b_+,\tau_0+t_g]$ constructively (destructively) interfere, such that we only have to consider the even time-derivatives of $s_I(t)$ on one of these intervals. We now obtain:

\begin{equation}
\begin{split}
    &\lambda \Big(s_I(b_-) - s_I(\tau_0)\Big) = \\
    &\sum_{k=0}^\infty \left\{\left(\frac{1}{2\omega_d}\right)^{2k+2}\lambda\gamma_1(2k+1,\beta)\left[\frac{\partial^{2k+2}s_I(t_1)}{\partial t_1^{2k+2}}\right]_{t_1=\tau_0}^{b_-} + \left(\frac{1}{2\omega_d}\right)^{2k+1}\gamma_1(2k+1,\beta)\left[\frac{\partial^{2k}s_I(t_1)}{\partial t_1^{2k}}\right]_{t_1=\tau_0}^{b_-}  \right\}.
\end{split}
\end{equation}

\noindent Similarly as for Eq. \eqref{eq:beta0_lambda_req}, if we set $\lambda=1/2\omega_d$, all terms in the sum cancel against each other except for the second term at $k=0$, which cancels against the term on the first line. To show that we additionally require $\Omega_I=\Omega_{I,\text{RWA}}$ on all three intervals we can readily use Eq. \eqref{eq:beta0_omega_req}. We immediately see that Eq. \eqref{eq:beta0_omega_req} is satisfied by substituting the bounds of the three intervals, adding the terms together, employing the symmetry of $s_I(t)$ around $t=t_g/2$ and by using that $s_I(\tau_0)=s_I(\tau_0+t_g)=0$.

Finally, we comment on an important difference between the analytical solutions for $\beta=0$ and symmetric integration windows. For the analytical solutions corresponding to $\beta=0$, the RWA and non-RWA time evolutions intersect at $t=t_0+nt_c$ for each $n$ and also at the end of the gate, as also apparent from Fig. \ref{fig:figure12}(d). For the analytical solutions corresponding to symmetric integration windows this is not the case, and the only guarantee is that the time evolutions intersect at the end of the gate. This difference in behavior arises from the fact that for $\beta=0$ the non-RWA terms in the time evolution are actually corrected \textit{in real time} by the pulse parameters. However, for the symmetric integration windows this is not the case, as the asymmetric terms in the time interval $t\in[\tau_0,b_-]$ cancel against those in the time interval $t \in [b_+,\tau_0+t_g]$. 

\section{First-order Magnus expansion}
\label{app:first-order-magnus-expansion}
In this section, we calculate the pulse parameters in the first-order Magnus approximation. This involves solving slightly more complicated expressions compared to the zeroth-order Magnus approximation. For example, for the detuning $\Delta$ we need to solve:

\begin{equation}
\label{eq:magnus1-detuning-requirement}
\begin{split}
    \frac{\Delta}{2}mt_c\sigma_z &= -\frac{i}{2}\int_{b_-}^{b_+}dt_1\int_{b_-}^{t_1}dt_1 \Big[A_I(t_1)\sigma_x+A_Q(t_1)\sigma_y, A_I(t_2)\sigma_x+A_Q(t_2)\sigma_y\Big] \\
    &=\int_{b_-}^{b_+}dt_1\int_{b_-}^{t_1}dt_1 \Big(A_I(t_1)A_Q(t_2) - A_Q(t_1)A_I(t_2)\Big)\sigma_z.
\end{split}
\end{equation}

\noindent Here, we again use general bounds $b_-=t_0+(n-m)t_c$ and $b_+=t_0+nt_c$. The double integrals in Eq. \eqref{eq:magnus1-detuning-requirement} are computed using the same approach as in the zeroth-order Magnus approximation in order to obtain expressions for the integrals as infinite series. Importantly, the terms in these sums now scale with $(2/N_c)^k$ compared to $(1/N_c)^k$ in the zeroth-order Magnus terms. This change in scaling arises from terms such as $\mathcal{E}_I^2(t)\sim \cos(4\pi t/t_g)$ which oscillate twice as fast as the original pulse envelope terms. Consequently, the pulse parameters can only be calculated for $N_c>2$, as for $N_c \leq 2$ the terms in these sums do not converge. This logic extends to higher order Magnus terms, since the $n$-th order Magnus term contains terms such as $\mathcal{E}_I^n(t) \propto \cos(2(n+1)\pi t/t_g)$. Hence, to derive the pulse parameters in the $n-$th order Magnus approximation, we require $N_c>n+1$ to ensure all the infinite series converge. This does not imply that solutions for the ideal pulse parameters do not exist in the $n$-th order Magnus approximation for $N_c \leq n+1$. For example, it is possible to evaluate the integrals in Eq. \eqref{eq:magnus1-detuning-requirement} for $N_c < 2$ using a similar approach but by swapping the roles of the oscillating terms and pulse envelopes in integration-by-parts such that the terms in the infinite series scale with $(N_C/2)^k$.

To compute the first-order correction terms, we make use of the following general solutions to the double integrals in Eq. \eqref{eq:magnus1-detuning-requirement}:

\begin{subequations}
\label{eq:very_scary_integrals}
\begin{flalign}
&\begin{aligned}
    & \int_{b_-}^{b_+}dt_1 f(t_1) \int_{b_-}^{t_1} dt_2 g(t_2) \sin(2\omega_dt_2 + 2\phi) = \sum_{k_1=0}^\infty \left(\frac{1}{2\omega_d}\right)^{k_1+1}\gamma_2(k_1,\beta) \frac{\partial^{k_1}g(t)}{\partial t^{k_1}}\bigg\rvert_{t=b_-} \int_{b_-}^{b_+} dt_1 f(t_1) - \\
    &\sum_{k_1=0}^\infty \sum_{k_2=0}^\infty \left(\frac{1}{2\omega_d}\right)^{2+k_1+k_2}(-1)^{\floor{\frac{k_1+2}{2}}} \Big(\chi_+(k_1)\gamma_1(k_2,\beta) - \chi_-(k_1)\gamma_2(k_2,\beta)\Big) \left[\frac{\partial^{k_2}}{\partial t^{k_2}} \left(f(t)\frac{\partial^{k_1} g(t)}{\partial t^{k_1}}\right) \right]_{t=b_-}^{b_+}
\end{aligned}&
\end{flalign}

\begin{flalign}
&\begin{aligned}
    & \int_{b_-}^{b_+}dt_1 f(t_1) \int_{b_-}^{t_1} dt_2 g(t_2) \cos(2\omega_dt_2 + 2\phi) = -\sum_{k_1=0}^\infty \left(\frac{1}{2\omega_d}\right)^{k_1+1}\gamma_1(k_1,\beta) \frac{\partial^{k_1}g(t)}{\partial t^{k_1}}\bigg\rvert_{t=b_-} \int_{b_-}^{b_+} dt_1 f(t_1) + \\
    &\sum_{k_1=0}^\infty \sum_{k_2=0}^\infty \left(\frac{1}{2\omega_d}\right)^{2+k_1+k_2}(-1)^{\floor{\frac{k_1}{2}}} \Big(\chi_-(k_1)\gamma_1(k_2,\beta) - \chi_+(k_1)\gamma_2(k_2,\beta)\Big) \left[\frac{\partial^{k_2}}{\partial t^{k_2}} \left(f(t)\frac{\partial^{k_1} g(t)}{\partial t^{k_1}}\right) \right]_{t=b_-}^{b_+}
\end{aligned}&
\end{flalign}

\begin{flalign}
&\begin{aligned}
    & \int_{b_-}^{b_+}dt_1 g(t_1)\sin(2\omega_dt_1 + 2\phi) \int_{b_-}^{t_1} dt_2 f(t_2) = \\
    &-\sum_{k=0}^\infty \left(\frac{1}{2\omega_d}\right)^{k+1}\gamma_2(k,\beta) \left( \left[\frac{\partial^k}{\partial t^k} \Big(g(t)f^{(1)}(t) \Big)\right]_{t=b_-}^{b_+} - f^{(1)}(b_-)\left[\frac{\partial^k g(t)}{dt^k} \right]_{t=b_-}^{b_+} \right)
\end{aligned}&
\end{flalign}

\begin{flalign}
&\begin{aligned}
    & \int_{b_-}^{b_+}dt_1 g(t_1)\cos(2\omega_dt_1 + 2\phi) \int_{b_-}^{t_1} dt_2 f(t_2) = \\
    &\sum_{k=0}^\infty \left(\frac{1}{2\omega_d}\right)^{k+1}\gamma_1(k,\beta) \left( \left[\frac{\partial^k}{\partial t^k} \Big(g(t)f^{(1)}(t) \Big)\right]_{t=b_-}^{b_+} - f^{(1)}(b_-)\left[\frac{\partial^k g(t)}{dt^k} \right]_{t=b_-}^{b_+} \right)
\end{aligned}&
\end{flalign}

\begin{flalign}
&\begin{aligned}
    & \int_{b_-}^{b_+}dt_1 \int_{b_-}^{t_1} dt_2 f(t_1)\sin(2\omega_dt_1+2\phi)g(t_2)\sin(2\omega_dt_2+2\phi) = \sum_{k=0}^\infty -\frac{(-1)^{\floor{\frac{k-1}{2}}}}{2}\left(\frac{1}{2\omega_d}\right)^{k+1}  \chi_{-}(k) \int_{b_-}^{b_+}dt f(t)\frac{\partial^k g(t)}{\partial t^k} + \\
    & \sum_{k_1=0}^\infty \sum_{k_2=0}^\infty \left(\frac{1}{2\omega_d}\right)^{k_1+k_2+2} \Bigg\{\left(\frac{1}{2}\right)^{k_1+2} \Big(\chi_-(k_2)\gamma_1(k_1,2\beta) + \chi_+(k_2)\gamma_2(k_1,2\beta)\Big) \left[\frac{\partial^{k_1}}{\partial t^{k_1}}\left(f(t)\frac{\partial^{k_2}g(t)}{\partial t^{k_2}}\right)\right]_{t=b_-}^{b_+} - \\
    &\hspace{4.3cm} \gamma_2(k_1,\beta)\gamma_2(k_2,\beta) \frac{\partial^{k_2}g(t)}{\partial t^{k_2}}\bigg\rvert_{t=b_-} \left[\frac{\partial^{k_1}f(t)}{\partial t^{k_1}}\right]_{t=b_-}^{b_+} \Bigg\}
\end{aligned}&
\end{flalign}

\begin{flalign}
&\begin{aligned}
    & \int_{b_-}^{b_+}dt_1 \int_{b_-}^{t_1} dt_2 f(t_1)\sin(2\omega_dt_1+2\phi)g(t_2)\cos(2\omega_dt_2+2\phi) = \sum_{k=0}^\infty\frac{(-1)^{\floor{\frac{k}{2}}}}{2}\left(\frac{1}{2\omega_d}\right)^{k+1}  \chi_{+}(k) \int_{b_-}^{b_+}dt f(t)\frac{\partial^k g(t)}{\partial t^k} + \\
    & \sum_{k_1=0}^\infty \sum_{k_2=0}^\infty \left(\frac{1}{2\omega_d}\right)^{k_1+k_2+2} \Bigg\{\left(\frac{1}{2}\right)^{k_1+2} \Big(-\chi_+(k_2)\gamma_1(k_1,2\beta) - \chi_-(k_2)\gamma_2(k_1,2\beta)\Big) \left[\frac{\partial^{k_1}}{\partial t^{k_1}}\left(f(t)\frac{\partial^{k_2}g(t)}{\partial t^{k_2}}\right)\right]_{t=b_-}^{b_+} + \\
    &\hspace{4.3cm} \gamma_2(k_1,\beta)\gamma_1(k_2,\beta) \frac{\partial^{k_2}g(t)}{\partial t^{k_2}}\bigg\rvert_{t=b_-} \left[\frac{\partial^{k_1}f(t)}{\partial t^{k_1}}\right]_{t=b_-}^{b_+} \Bigg\}
\end{aligned}&
\end{flalign}

\begin{flalign}
&\begin{aligned}
    & \int_{b_-}^{b_+}dt_1 \int_{b_-}^{t_1} dt_2 f(t_1)\cos(2\omega_dt_1+2\phi)g(t_2)\sin(2\omega_dt_2+2\phi) = \sum_{k=0}^\infty -\frac{(-1)^{\floor{\frac{k-1}{2}}}}{2}\left(\frac{1}{2\omega_d}\right)^{k+1}  \chi_{+}(k) \int_{b_-}^{b_+}dt f(t)\frac{\partial^k g(t)}{\partial t^k} + \\
    & \sum_{k_1=0}^\infty \sum_{k_2=0}^\infty \left(\frac{1}{2\omega_d}\right)^{k_1+k_2+2} \Bigg\{\left(\frac{1}{2}\right)^{k_1+2} \Big(-\chi_+(k_2)\gamma_1(k_1,2\beta) + \chi_-(k_2)\gamma_2(k_1,2\beta)\Big) \left[\frac{\partial^{k_1}}{\partial t^{k_1}}\left(f(t)\frac{\partial^{k_2}g(t)}{\partial t^{k_2}}\right)\right]_{t=b_-}^{b_+} + \\
    &\hspace{4.3cm} \gamma_1(k_1,\beta)\gamma_2(k_2,\beta) \frac{\partial^{k_2}g(t)}{\partial t^{k_2}}\bigg\rvert_{t=b_-} \left[\frac{\partial^{k_1}f(t)}{\partial t^{k_1}}\right]_{t=b_-}^{b_+} \Bigg\}
\end{aligned}&
\end{flalign}

\begin{flalign}
&\begin{aligned}
    & \int_{b_-}^{b_+}dt_1 \int_{b_-}^{t_1} dt_2 f(t_1)\cos(2\omega_dt_1+2\phi)g(t_2)\cos(2\omega_dt_2+2\phi) = \sum_{k=0}^\infty \frac{(-1)^{\floor{\frac{k}{2}}}}{2}\left(\frac{1}{2\omega_d}\right)^{k+1}  \chi_{-}(k) \int_{b_-}^{b_+}dt f(t)\frac{\partial^k g(t)}{\partial t^k} + \\
    & \sum_{k_1=0}^\infty \sum_{k_2=0}^\infty \left(\frac{1}{2\omega_d}\right)^{k_1+k_2+2} \Bigg\{\left(\frac{1}{2}\right)^{k_1+2} \Big(\chi_-(k_2)\gamma_1(k_1,2\beta) - \chi_+(k_2)\gamma_2(k_1,2\beta)\Big) \left[\frac{\partial^{k_1}}{\partial t^{k_1}}\left(f(t)\frac{\partial^{k_2}g(t)}{\partial t^{k_2}}\right)\right]_{t=b_-}^{b_+} - \\
    &\hspace{4.3cm} \gamma_1(k_1,\beta)\gamma_1(k_2,\beta) \frac{\partial^{k_2}g(t)}{\partial t^{k_2}}\bigg\rvert_{t=b_-} \left[\frac{\partial^{k_1}f(t)}{\partial t^{k_1}}\right]_{t=b_-}^{b_+} \Bigg\}
\end{aligned}&
\end{flalign}

\end{subequations}

\noindent Here, $g^{(1)}(t)$ denotes the first antiderivative of $g(t)$. Solving for the ideal pulse parameters using these general solutions is still challenging, since Eq. \eqref{eq:very_scary_integrals} together with Eq. \eqref{eq:magnus0-requirement} form a very nonlinear system of equations. For example, the integrals in Eq. \eqref{eq:very_scary_integrals} define the detuning, but they themselves depend on the detuning since the integration windows depend on $\omega_d$ as well as $\Omega_I$ and $\lambda$ which depend on $\omega_d$ in a similar way. Therefore, we use fixed-point iteration to obtain the pulse parameters, which we find to converge well.

After solving and numerically verifying the correction terms in the first-order Magnus approximation, we found the calculated pulse parameters to be highly inaccurate. This is caused by non-negligible effects from terms outside the integration windows, i.e. with $t\in [0,t_0] \cup [t_0+N_ct_c,t_g]$. We find this effect to be most prominent in the first-order corrections to $\lambda$. Consider, for example, the term:

\begin{equation}
    \zeta_1(b_-,b_+) = \frac{\Delta}{2}\int_{b_-}^{b_+}dt_1 \int_{b_-}^{t_1}dt_2 \Big(\mathcal{E}_I(t_1) - \mathcal{E}_I(t_2)\Big).
\end{equation}

\noindent This term integrates exactly to $0$ over the entire duration of the gate, i.e. $\zeta_1(0,t_g)=0$, but is non-zero and on the same order as the correction terms in the zeroth-order Magnus approximation for other values of $b_-$ and $b_+$. This severely compromises the accuracy of the calculated ideal pulse parameters, as we are effectively correcting for a term that trivially integrates to 0. To resolve this issue, we split the carrier signal into two terms: one term that is commensurate with the gate duration and a term that is incommensurate. We then absorb the incommensurate term into the pulse envelopes and define the commensurate term as the new carrier signal. This enables integrating over the full time evolution for arbitrary gate durations and drive frequencies at the cost of slower convergence of the infinite series expressions for the pulse parameters. Concretely, we rewrite:

\begin{equation}
\begin{split}
    & \mathcal{E}(t)\cos(2\omega_dt+2\phi) = \mathcal{E}(t)\Big(\cos(2\hat{\omega}_dt + 2\hat
    \phi)\cos(2\tilde{\omega}_dt + 2\tilde{\phi}) - \sin(2\hat{\omega}_dt + 2\hat
    \phi)\sin(2\tilde{\omega}_dt + 2\tilde{\phi})\Big), \\
    & \mathcal{E}(t)\sin(2\omega_dt+2\phi) = \mathcal{E}(t)\Big(\sin(2\hat{\omega}_dt + 2\hat
    \phi)\cos(2\tilde{\omega}_dt + 2\tilde{\phi}) + \cos(2\hat{\omega}_dt + 2\hat
    \phi)\sin(2\tilde{\omega}_dt + 2\tilde{\phi})\Big).
\end{split}
\end{equation}

\noindent Here, the oscillating terms $\propto \hat{\omega}_d$ take the role of carrier signal, while we absorb the oscillating terms $\propto \tilde{\omega}_d$ into the pulse envelope, here denoted in a general form as $\mathcal{E}(t)$. We also redefine the carrier phase to ensure that the updated pulse envelopes maintain the (anti-)symmetric properties of their derivatives around $t=t_g/2$, although this is not strictly necessary. More specifically, we have $\hat{N}_c = \lfloor t_g\omega_d/\pi \rceil$, $\hat{\omega}_d=\pi\hat{N}_c/t_g$, $\tilde{\omega}_d=\omega_d - \hat{\omega}_d$, $\tilde{\phi}=-\tilde{\omega}_dt_g/2$ and $\hat{\phi}=\phi + \tilde{\omega}_dt_g/2$. Here, $\lfloor . \rceil$ denotes rounding to the nearest integer. Notice that $\abs{\tilde{\omega}_d} \leq \pi/2t_g$, such that the infinite series expressions will still converge as long as $N_c > 2$ for the zeroth-order Magnus approximation and $N_c > 3$ for the first-order Magnus approximation. The pulse parameters calculated in Fig. \ref{fig:figure3} are computed up to $15^\text{th}$ order in $3/N_c$. Using this substitution, we can force the carrier signal to be commensurate with the gate duration, such that we can always integrate over the full gate duration and we are no longer limited by Magnus intervals. While it may seem as if we effectively lose the $t_0$ degree of freedom here, this is not the case. This degree of freedom is expressed through the choice of dividing the total carrier phase $\phi$ between $\tilde{\phi}$ and $\hat{\phi}$. Here, this division is fixed to maintain the symmetry properties of the pulse envelopes. To compute the ideal pulse parameters, we need to minimize the gate error averaged over all carrier phases as a function of the pulse parameters, i.e. we need to solve:

\begin{equation}
    \min_{\mathcal{P}(\phi)} \mathbb{E}_\phi \Big[\text{Error}(\phi)\Big].
\end{equation}

\noindent This is highly non-trivial, and we simply compute the ideal pulse parameters by averaging over the carrier phases:

\begin{equation}
    \mathcal{P} = \mathbb{E}_\phi \big[ \mathcal{P}(\phi) \big]
\end{equation}

\section{Modeling a strongly anharmonic system as a two-level system \label{app:fluxonium_as_tls}}
Modeling a strongly anharmonic system, such as the fluxonium, as a two-level system requires the derivation of an effective Hamiltonian in the computational subspace of the four-level Hamiltonian given in Eq. \eqref{eq:drive_hamiltonian}. We perform an adiabatic elimination using:

\begin{equation}
\begin{split}
    & \tilde{H}_\text{eff}(t) = e^{-iS(t)}\tilde{H}(t)e^{iS(t)} + \frac{\partial S(t)}{\partial t} = \tilde{H}(t) - i\Big[S(t),\tilde{H}(t)\Big] + \frac{1}{2}\Big[S(t),\Big[S(t),\tilde{H}(t)\Big]\Big] + \dots + \frac{\partial S(t)}{\partial t}, \\
    & S(t) = -\frac{\eta_{12}A_y(t)}{\alpha_2}\sigma_x^{12} + \frac{\eta_{12}A_x(t)}{\alpha_3}\sigma_y^{12} - \frac{\eta_{03}A_y^{03}(t)}{\alpha_3}\sigma_x^{03} + \frac{\eta_{03}A_x^{03}(t)}{\alpha_3}\sigma_y^{03}.
\end{split}
\end{equation}

\noindent Neglecting terms $\mathcal{O}(S^2)$ and higher, the effective Hamiltonian becomes:

\begin{equation}
\label{eq:adiabatic_elimination_final_delta}
\begin{split}
    & \tilde{H}_\text{eff}(t) = \big(\omega_1 - \omega_d' + \Delta'(t)\big)\ket{1}\bra{1} + \sum_{jk \in \{01,02,13\} } A_x^{jk}(t)\eta_{ij}\sigma_x^{jk} + A_y^{jk}(t)\eta_{jk}\sigma_y^{jk}, \\
    & \Delta'(t) = \frac{\mathcal{E}_I^2(t)+\mathcal{E}_Q^2(t)}{2}\left(\frac{\eta_{03}^2}{\alpha_3} - \frac{\eta_{12}^2}{\alpha_2}\right),
\end{split}
\end{equation}

\begin{figure}[t]
\centering
\includegraphics[width=\linewidth]{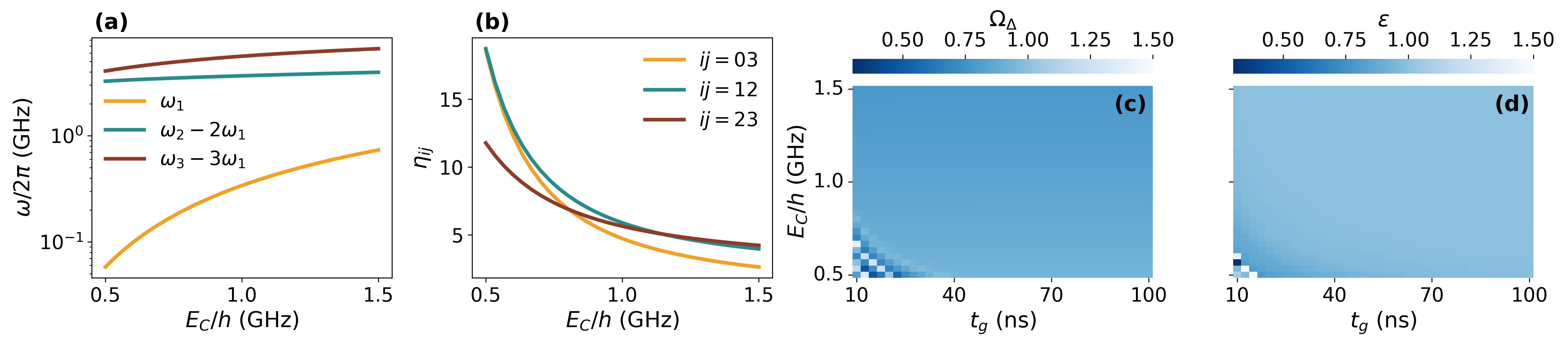}
\caption{\label{fig:figure5-appedix} Fluxonium parameters and pulse parameters for the results shown in Fig. \ref{fig:figure5}. (a) and (b) show the energy levels and relative drive strengths as a function of $E_C$ respectively. (c) and (d) show the optimized parameters $\Omega_\Delta$ and $\epsilon$ respectively.}
\end{figure}

\noindent where we have neglected all non-stationary terms in $\Delta'(t)$. Notice that $\Delta'(t)$ in Eq. \eqref{eq:adiabatic_elimination_final_delta} is equal to Eq. \eqref{eq:time_dependent_detuning} up to a change in sign. Here, we have essentially solved the opposite problem compared to the main text: we calculated how the drive terms in the two-level Hamiltonian change due to the effect from higher levels. However, we are interested in the opposite problem: how do we need to change the pulse parameters in the four-level Hamiltonian such that we can model the system by the two-level Hamiltonian in Eq. \eqref{eq:hamiltonian-terms}, explaining the change in sign.

Notice that, due to the strong coupling of the higher order transitions in the fluxonium qubit, the adiabaticity of this transformation is compromised, as $|\mathcal{E}_I(t)\eta_j/\alpha_j|$ can approach the order of unity. This explains why the error in Fig. \ref{fig:figure5}(b) is large for short gate durations and heavy fluxonium parameters.

In the main text we optimized $\Omega_\Delta$ and $\epsilon$ such that we can model the fluxonium by a two-level system as a function of the gate duration and $E_C$. In Fig. \ref{fig:figure5-appedix}(a) and (b) we plot the fluxonium parameters belonging to this range of $E_C$'s. In Fig. \ref{fig:figure5-appedix}(c) and (d) we plot the fitted parameters $\Omega_\Delta$ and $\epsilon$. We see that, for the majority of the parameter range, the fitted parameters are close to 1, showing that the adiabatic transformation derived in this section effectively captures the effect from higher order levels for this parameter range.

\vspace{0.6cm}
\twocolumngrid
\section{Device and experimental setup}
\label{app:device_and_setup}
\begin{figure}[t]
\centering
\includegraphics[width=\linewidth]{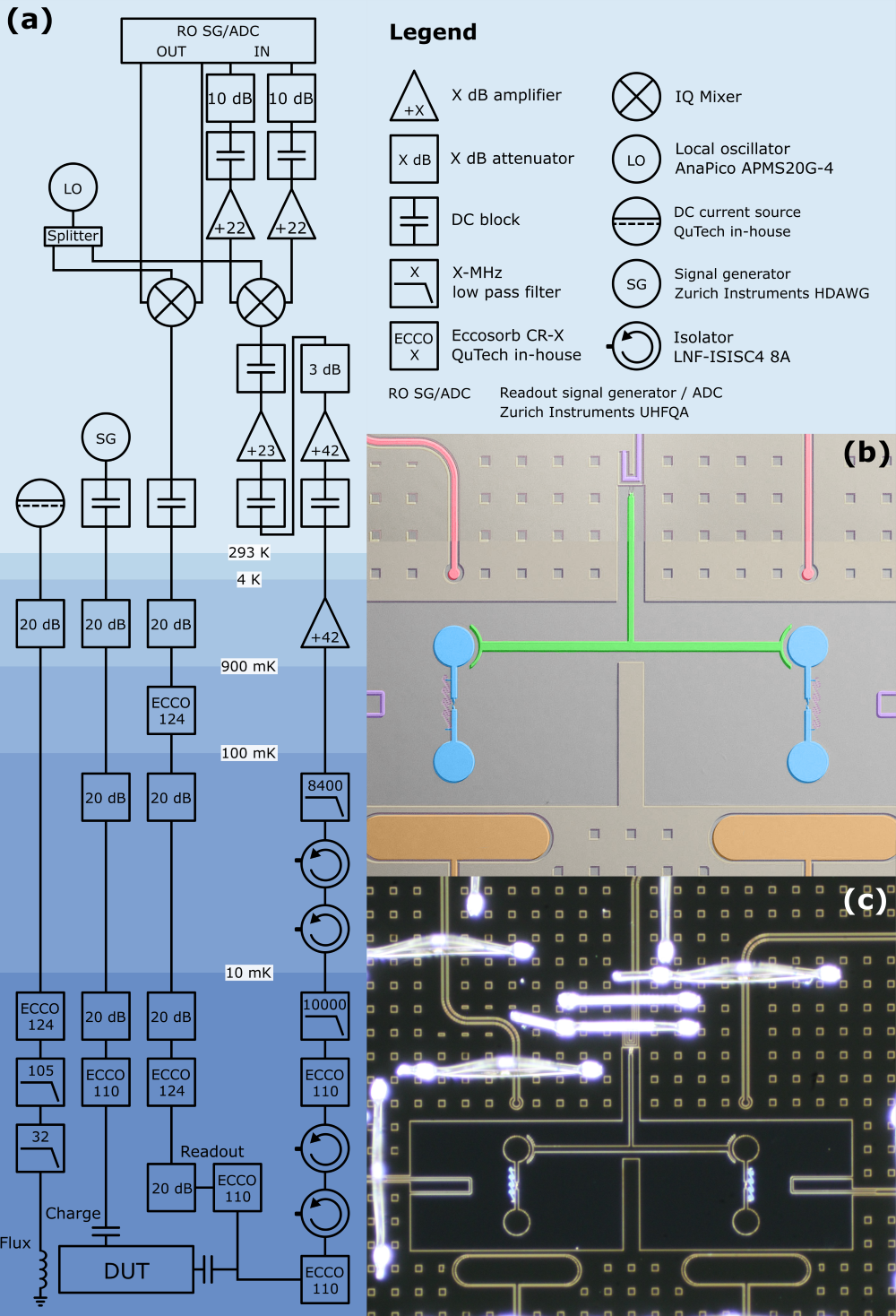}
\caption{\label{fig:experimental_setup} (a) Schematic of the electronic setup used in the experiments. (b) False-colored image of the device. The two fluxonium qubits are shown in blue, and the tunable transmon in green. The charge and flux lines are shown in pink and purple respectively, and the pad of the readout resonator is shown in brown. (c) Optical microscope image of the fabricated device.}
\end{figure}

The fluxonium qubit on which the experiments were performed was part of a two-fluxonium system in which the fluxoniums are coupled by a tunable transmon. We experimentally obtained the fluxonium's energy parameters as $E_C/h=0.88$ GHz, $E_L/h=0.50$ GHz and $E_J/h=4.92$ GHz. The fluxonium is capacitively coupled to a 4.993 GHz resonator used for dispersive readout. For more details on the device, we refer the reader to Ref. \cite{sidd_eugene}. A false-colored image of the device is shown in Fig. \ref{fig:experimental_setup}(b) and an optical microscope image is shown in Fig. \ref{fig:experimental_setup}(c). During these experiments, both fluxoniums were parked at $\varphi_\text{ext}=0.5$ and the transmon was at a flux bias point that suppressed the residual $zz$ interaction between the fluxoniums. Specifically, we measure a residual $zz$ interaction of $0.5 \pm 2$ kHz. 

The device was cooled down in a Bluefors LD400 dilution refrigerator to a base temperature of 8 mK. The device is protected from electromagnetic noise by an aluminum shield that is placed directly on top of the PCB. The device is further protected from thermal and electromagnetic noise by one copper can and two MuMETAL cans. The full electronic setup is displayed in Fig. \ref{fig:experimental_setup}(a). To flux-bias the qubits, we use an in-house made DC current generator. The qubit drive pulses were generated by a Zurich Instruments HDAWG at a sampling rate of 2.4 GS/s. Given the low frequency of the fluxonium qubit, no upconversion of the drive pulses is required. Due to the challenges with microwave heating in this experiment, we do not use an active reset such as a sideband reset \cite{fluxonium_initialization}. Instead, we make use of post selection, in which each circuit is preceded by a measurement to determine the state of the qubit at the start of the circuit. To counteract heating from the drive pulses, this initial measurement is preceded by a delay time of $700$ $\mu$s. There is another $7$ $\mu$s delay between the initial measurement and the start of the actual circuit to ensure the readout resonator is depleted again. The readout pulses are generated and analyzed by the Zurich Instruments UHFQA. They are upconverted using the Zurich Instruments HDIQ mixer with an LO signal from the AnaPico APMS20G-4. All signals pass through a series of filters, attenuators, and in-house made Eccosorb IR filters. The output signal passes through a series of cryogenic dual-junction isolators, a cryogenic HEMT (LNF-LNC4\_8C), a room temperature HEMT (LNF-LNR4\_8ART), and through a 23 dB amplifier (Mini-Circuits ZRON-8G+) before being demodulated and passing through one more amplifier (Mini-Circuits GALI-3+).

\section{Extended data}
\label{app:extended_data}

\subsection{Heating and incoherent errors}
\begin{figure}[t]
\centering
\includegraphics[width=\linewidth]{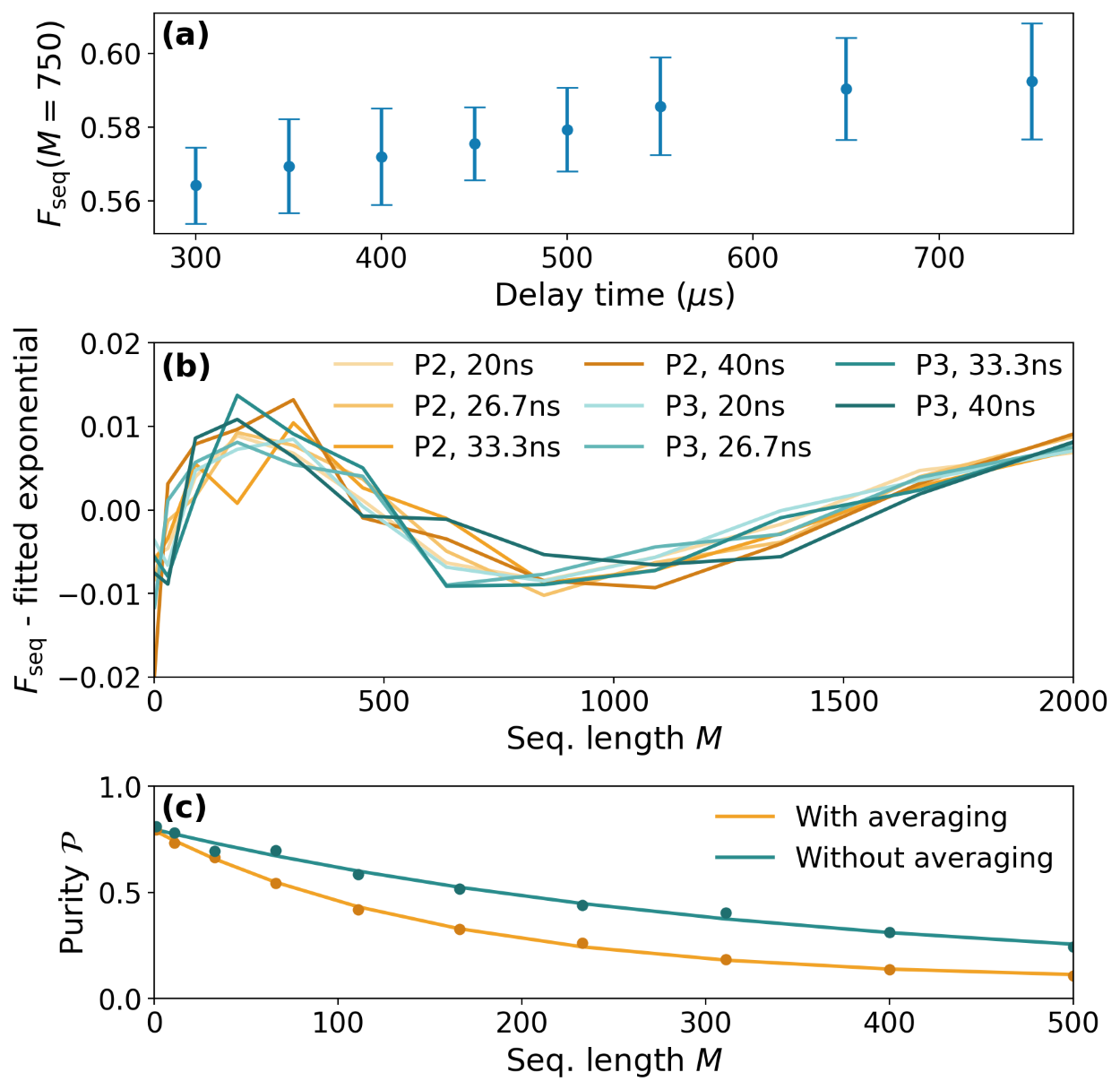}
\caption{\label{fig:rb_exp_appendix} (a) RB sequence fidelity for $M=750$ and $t_g=20$ ns as a function of the initial delay time. (b) Difference between the measured sequence fidelities and the fitted exponential decay curve for all RB measurements of protocol P2 and P3 with $t_g\geq 20$ ns. (c) Purity RB decay curves for $t_g=13.3$ ns with and without averaging each circuit over the carrier phase. The markers indicate the measured data, and the solid line the fitted exponential decay curve.}
\end{figure}

Here, we elaborate on the discrepancy between the incoherent error rate measured using PRB and the expected incoherent error from decoherence processes. For $t_g\geq 20$ ns we suspect that the majority of the discrepancy arises from heating from the microwave pulses. To verify this, we measure the RB sequence fidelity for a sequence length of $M=750$ and for $t_g=20$ ns as a function of the delay time before the post selection measurement, of which the results are shown in Fig. \ref{fig:rb_exp_appendix}(a). We see that the sequence fidelity improves by increasing the delay time, which is very typical for microwave heating. For the experiments presented in this work the delay time was fixed to $700$ $\mu$s. However, we found that significant effects from heating remained. To highlight this, we plot the difference between the measured RB sequence fidelities and the fitted exponential decay curve for the RB experiments performed for protocol P2 and P3 and for all gate durations $t_g \geq 20$ ns in Fig. \ref{fig:rb_exp_appendix}(b). We see that the data consistently deviates from an exponential decay curve. The sequence fidelities for short sequence lengths are higher than the fitted exponential decay, whereas the sequence fidelity for longer sequences is lower than the fitted exponential decay. This is an indication that there is still significant heating from the microwave pulses that increases the error rate as a function of the sequence length. 

For the $13.3$ ns gate, the discrepancy between the incoherent error measured using PRB and estimated from decoherence is much higher than for the other gate durations. We ascribe this to the increased dependence of the time evolution on the carrier phase. PRB essentially measures the decay of the variance of the sequence fidelities of random Clifford circuits. It takes advantage of the fact that incoherent errors reduce this variance, while coherent errors do not. For an increasing amount of non-RWA errors, the dependence of the time evolution on the carrier phase also increases. We suspect that, since we average each circuit over the carrier phase by incrementing the carrier phase by 1 degree for each repetition of the circuit, this increasing dependence on the carrier phase amounts to an increase in the decay of the variance of the sequence fidelities. To verify this, we measure PRB for a 13.3 ns gate in Fig. \ref{fig:rb_exp_appendix}(c) with and without averaging over the carrier phases. Specifically, we plot the normalized purity $\mathcal{P}=\sqrt{\langle \sigma_x \rangle^2 + \langle \sigma_y \rangle^2 + \langle \sigma_z \rangle^2}$ versus the sequence length $M$. If we average over the carrier phase, we measure an incoherent error rate of $1.7\cdot 10^{-3}$. By disabling this averaging and measuring each circuit only for one specific carrier phase the measured incoherent error rate improves significantly to $0.8 \cdot 10^{-3}$. We conclude that the increase in the variance of the coherent error compromises the accuracy of the incoherent error rate measured using PRB.

\subsection{Phase error heatmaps}
\begin{figure}[t]
\centering
\includegraphics[width=\linewidth]{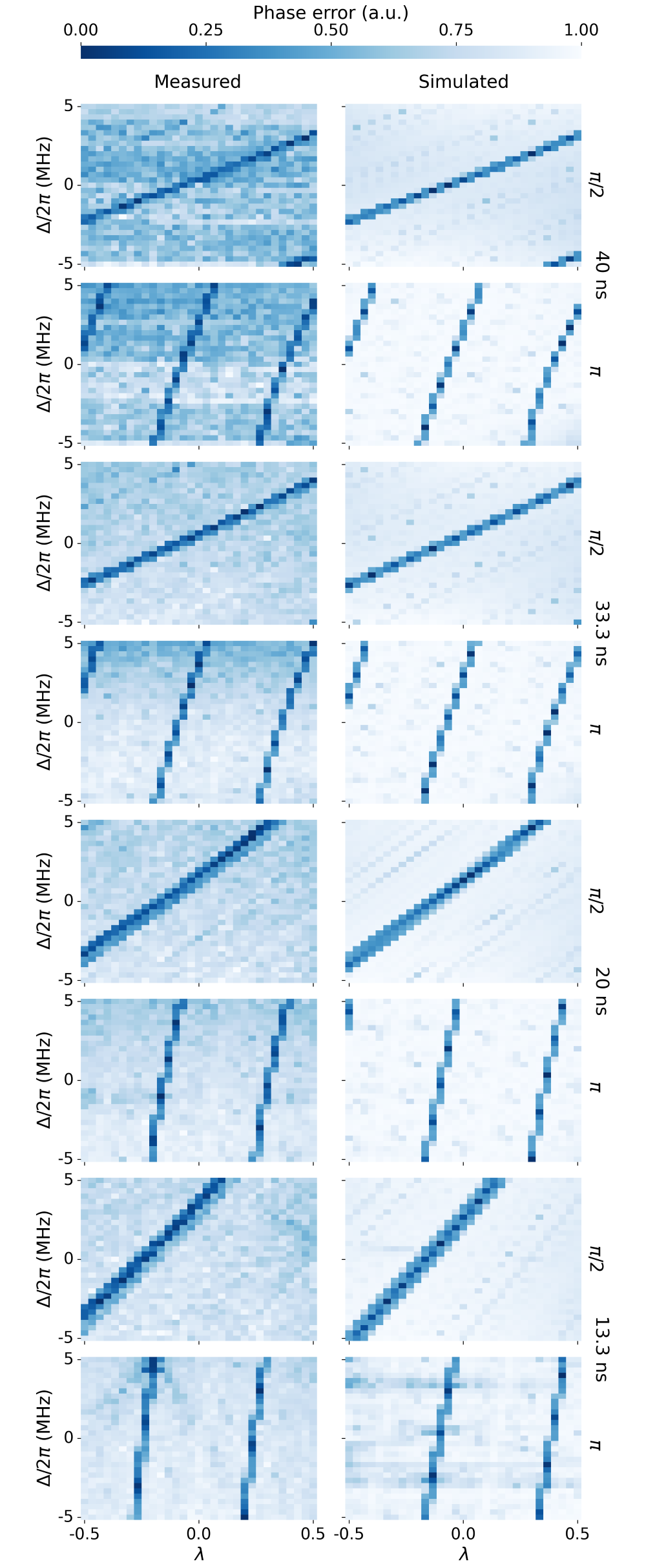}
\caption{\label{fig:all_heatmaps} All measured and simulated heatmaps of the phase error as a function of the PPP $\lambda$ and the detuning $\Delta$.}
\end{figure}

In Fig. \ref{fig:all_heatmaps} we plot the phase error heatmaps for the remaining gate durations. We plot heatmaps for $\pi$-rotations as well as for $\pi/2$ rotations. We further verified these heatmaps by simulating the experiments numerically. We find very good correspondence between the experimental and numerical heatmaps. Simulating the phase error pseudo-identity circuits raises a computational challenge. If the RWA holds, each operation only needs to be solved once for a specified value of $\lambda$ and $\Delta$. Here, due to the dependence of the time evolution on the carrier phase, every operation in a circuit is different and needs to be simulated separately. Since we additionally average the experimentally executed circuits over the initial carrier phase, simulating these experiments is a computationally expensive task. Therefore, for each gate duration, we numerically solve the four-level time evolution for a set range of 41 values for the detuning $\Delta$, PPP $\lambda$ and carrier phase $\phi$ each. We then truncate the time evolution to the qubit subspace and fit the rotation angle and axis. We linearly interpolate those values to obtain the time evolution for arbitrary values of $\Delta$, $\lambda$ and $\phi$. This significantly reduces the total number of time evolutions that need to be solved to approximately 1.4 million, for which we make use of a HPC \cite{DHPC2024}. 

\subsection{Error budgets}
\begin{figure}[t]
\centering
\includegraphics[width=\linewidth]{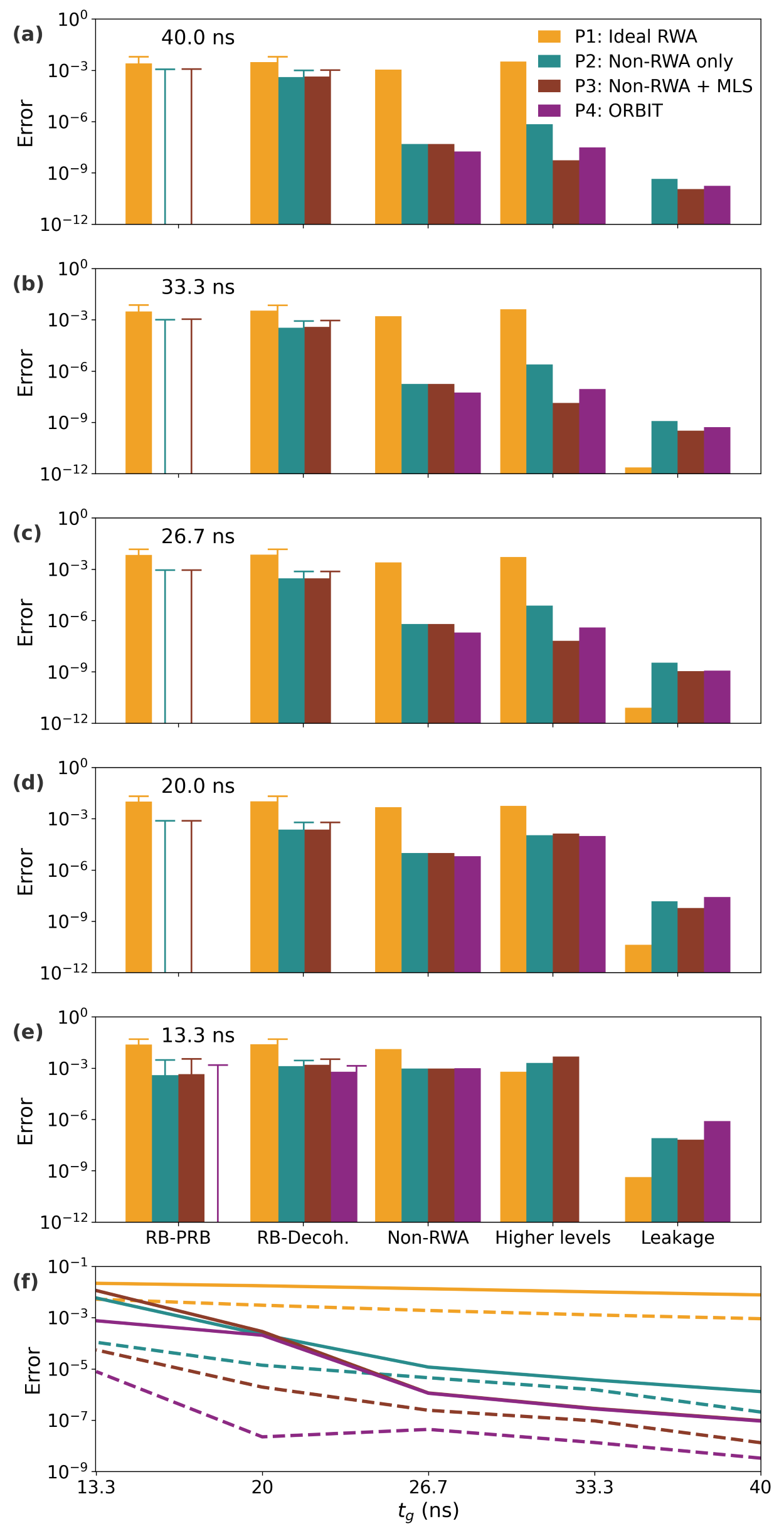}
\caption{\label{fig:error_budget} (a)-(e) Error budgets for all gate durations. \mbox{(f) Total} estimated coherent error for the $\pi/2$ gates (dashed lines) and $\pi$ gates (solid lines).}
\end{figure}

Fig. \ref{fig:error_budget}(a)-(e) shows the error budgets for all protocols and all gate durations. The procedure for computing the error budgets is outlined in the main text. Fig. \ref{fig:error_budget}(f) shows the total estimated coherent error for $\pi$ and $\pi/2$ gates individually. Depending on the gate duration and calibration protocol, the $\pi/2$ gates achieve coherent errors that are $1-4$ order of magnitude smaller than the $\pi$ gates. This is ascribed to the exponential increase in the coherent error when the RWA becomes increasingly invalid. Therefore, it is possible to achieve lower coherent errors for shorter gate durations by using a universal gate set that is made-up of only $\pi/2$ gates. Finally, Fig. \ref{fig:error_budget} shows that the deterministic calibration protocols P2 and P3 achieve error rates averaged over the $\pi$ and $\pi/2$ gates that are typically within 1.5 and 1 order of magnitude of the errors computed with protocol P4 respectively. This shows that the deterministic calibration protocols are very effective at achieving error rates that are close to the minimum achievable error using the pulse shapes developed in this work.

\FloatBarrier

\bibliography{apssamp}

\end{document}